\documentclass[fleqn,usenatbib]{mnras}

\usepackage[T1]{fontenc}
\usepackage{graphicx}
\usepackage{amsmath}
\usepackage{amssymb}
\usepackage{verbatim}

\title[Ultra-massive quiescent galaxies at $3 < z < 5$]{The JWST EXCELS survey: Too much, too young, too fast? Ultra-massive quiescent galaxies at $\mathbf{3 < z < 5}$}
\author[A. C. Carnall et al.]
{A. C. Carnall$^{1}$\thanks{E-mail: adamc@roe.ac.uk},
F. Cullen$^{1}$,
R. J. McLure$^{1}$,
D. J. McLeod$^{1}$,
R. Begley$^{1}$,
C. T. Donnan$^{1}$,
\newauthor
J. S. Dunlop$^{1}$,
A. E. Shapley$^{2}$,
K. Rowlands$^{3, 4}$,
O. Almaini$^{5}$,
K. Z. Arellano-C\'{o}rdova$^{1}$,
\newauthor
L. Barrufet$^{1}$,
A. Cimatti$^{6, 7}$,
R. S. Ellis$^{8}$,
N. A. Grogin$^{9}$,
M. L. Hamadouche$^{1}$,
G. D. Illingworth$^{10}$,
\newauthor
A. M. Koekemoer$^{9}$,
H.-H. Leung$^{11}$,
C. C. Lovell$^{12}$,
P. G. P\'erez-Gonz\'alez$^{13}$,
P. Santini$^{14}$,
\newauthor
T. M. Stanton$^{1}$,
V. Wild$^{11}$
\medskip\\
$^{1}$ SUPA\thanks{Scottish Universities Physics Alliance}, Institute for Astronomy, University of Edinburgh, Royal Observatory, Edinburgh EH9 3HJ, UK \\
$^{2}$ Department of Physics \& Astronomy, University of California, Los Angeles, 430 Portola Plaza, Los Angeles, CA 90095, USA \\
$^{3}$ AURA for ESA, Space Telescope Science Institute, 3700 San Martin Drive, Baltimore, MD 21218, USA\\
$^{4}$ William H. Miller III Department of Physics and Astronomy, Johns Hopkins University, Baltimore, MD 21218, USA \\
$^{5}$ University of Nottingham, School of Physics and Astronomy, Nottingham, NG7 2RD, UK\\
$^{6}$ Department of Physics and Astronomy (DIFA), University of Bologna, Via Gobetti 93/2, I-40129, Bologna, Italy\\
$^{7}$ INAF, Osservatorio di Astrofisica e Scienza dello Spazio, Via Piero Gobetti 93/3, I-40129, Bologna, Italy\\
$^{8}$ Department of Physics and Astronomy, University College London, Gower Street, London WC1E 6BT, UK\\
$^{9}$ Space Telescope Science Institute, 3700 San Martin Drive, Baltimore, MD 21218, USA\\
$^{10}$ Department of Astronomy and Astrophysics, University of California, Santa Cruz, CA 95064, USA\\
$^{11}$ SUPA\footnotemark[2], School of Physics and Astronomy, University of St Andrews, North Haugh, St Andrews KY16 9SS, UK\\
$^{12}$ Institute of Cosmology and Gravitation, University of Portsmouth, Burnaby Road, Portsmouth, PO1 3FX, UK\\
$^{13}$ Centro de Astrobiolog\'ia (CAB), CSIC-INTA, Ctra. de Ajalvir km 4, Torrej\'on de Ardoz, E-28850, Madrid, Spain\\
$^{14}$ INAF, Osservatorio Astronomico di Roma, via di Frascati 33, 00078 Monte Porzio Catone (RM), Italy\vspace{-0.5cm}}

\date{Accepted XXX. Received YYY; in original form ZZZ\vspace{-0.3cm}}

\pubyear{2024}

\begin{document}
\label{firstpage}
\pagerange{\pageref{firstpage}--\pageref{lastpage}}
\maketitle

\begin{abstract}\vspace{-0.2cm}

\noindent We report ultra-deep, medium-resolution spectroscopic observations for 4 quiescent galaxies with log$_{10}(M_*/\mathrm{M_\odot})>11$ at $3 < z < 5$. These data were obtained with JWST NIRSpec as part of the Early eXtragalactic Continuum and Emission Line Science (EXCELS) survey, which we introduce in this work. The first two galaxies are newly selected from PRIMER UDS imaging, both at $z=4.62$ and separated by $860$ pkpc on the sky, within a larger structure for which we confirm several other members. Both formed at $z\simeq8-10$. These systems could plausibly merge by the present day to produce a local massive elliptical galaxy. The other two ultra-massive quiescent galaxies are previously known at $z=3.99$ and $3.19$, with the latter (ZF-UDS-7329) having been the subject of debate as potentially too old and too massive to be accommodated by the $\Lambda$-CDM halo-mass function. Both exhibit high stellar metallicities, and for ZF-UDS-7329 we are able to measure the $\alpha-$enhancement, obtaining [Mg/Fe]~=~$0.42^{+0.19}_{-0.17}$. We finally evaluate whether these 4 galaxies are consistent with the $\Lambda$-CDM halo-mass function using an extreme value statistics approach. We find that the $z=4.62$ objects and the $z=3.19$ object are unlikely within our area under the assumption of standard stellar fractions ($f_*\simeq0.1-0.2$). However, these objects roughly align with the most massive galaxies expected under the assumption of 100 per cent conversion of baryons to stars ($f_*$=1). Our results suggest extreme galaxy formation physics during the first billion years, but no conflict with $\Lambda$-CDM cosmology.\\
\end{abstract}

\begin{keywords}
galaxies: evolution - galaxies: formation - galaxies: high-redshift - galaxies: statistics - galaxies: stellar content\vspace{-0.35cm}
\end{keywords}

\section{Introduction} \label{sec:intro}

Since the launch of JWST, much attention has focused on the earliest stages of massive galaxy formation. In particular, many studies have reported candidate massive galaxies at high redshift that were too faint and/or too red to have been detected (or at least to have their redshifts measured) with previous instrumentation (e.g., \citealt{Labbe2023, Barrufet2023, Xiao2023, Gottumukkala2023, Weibel2024}). This has led to much discussion as to whether such objects are in fact too early, too massive and too numerous to be accommodated within our current understanding of galaxy formation physics, or even within the $\Lambda$-CDM cosmological framework (e.g., \citealt{Boylan-Kolchin2023, Chworowsky2023, Harvey2024}).

Another surprising finding from early JWST data has been the discovery that quenching in massive galaxies is far more common during the first two billion years ($z>3$) than previously suspected. This gives rise to higher-than-expected number densities for massive quiescent galaxies at $3 < z < 5$ (e.g., \citealt{Carnall2023b, Valentino2023, Long2023, Alberts2023}), with galaxy formation models already being updated in light of these results (e.g., \citealt{Lagos2023}). This approach of studying rare and extreme high-redshift massive galaxies has a long history of providing key constraints on contemporary models for galaxy formation and cosmology (e.g., \citealt{Dunlop1996, Cimatti2004, Fontana2009}). There is also interest in the extent to which the earliest massive galaxies go on to form the most massive galaxies in the cores of local galaxy clusters (e.g., see \citealt{Remus2023, Hartley2023, Rennehan2024}).

Aside from the obvious significant interest in how quenching can occur so early and so rapidly in so many galaxies, another interesting opportunity afforded by early quiescent galaxies is to trace back their star-formation histories (SFHs) to constrain when they first began forming stars. The recovery of SFHs is more tractable for quiescent galaxies than their star-forming counterparts, as quiescent galaxy spectra are far less affected by the outshining of older stellar populations by young stars. This kind of analysis is now greatly aided by the capability of JWST NIRSpec \citep{Jakobsen2022} to provide high signal-to-noise ratio (SNR) medium-resolution continuum spectroscopy for faint, red galaxies for the first time (e.g., \citealt{Nanayakkara2024, Setton2024, Barrufet2024, Slob2024, Park2024}), enabling the age-metallicity-dust degeneracy to be broken (e.g., \citealt{Conroy2013}).

In \cite{Carnall2023c}, we have reported the spectroscopic confirmation of the first massive quiescent galaxy significantly beyond $z=4$, GS-9209 at $z=4.658$, from JWST NIRSpec Cycle 1 data. We show via full spectral fitting that this galaxy formed its stellar mass of log$_{10}(M_*/\mathrm{M_\odot})=10.58\pm0.02$ over a $\simeq200$ Myr period from $\simeq600-800$ Myr after the Big Bang, before quenching at $z_\mathrm{quench}=6.5^{+0.2}_{-0.5}$. 

We also uncover several indicators of active galactic nucleus (AGN) activity in this galaxy, most notably extremely broad H$\alpha$ emission. Based on this broad line, we infer a central black hole mass of $M_\bullet\simeq0.5-1.0\times10^9\ \mathrm{M_\odot}$, which places GS-9209 approximately a factor of 5 above the local relation between galaxy stellar mass and black hole mass. This result is consistent with previous work tracing the evolution of this relationship from the local Universe towards cosmic noon \citep{McLure2006}, local analogues (e.g., \citealt{Ferre-Mateu2015}), and also with recent predictions from the \textsc{Simba-C} cosmological simulation \citep{Szpila2024}.

For its central black hole to have become so massive, this galaxy must have experienced substantial AGN activity, suggesting quasar-mode AGN feedback as a likely mechanism responsible for quenching its star formation (e.g., \citealt{Maiolino2012, Hartley2023, Kimmig2023, Belli2023}). However, as high-SNR medium-resolution spectroscopy is only available for a single massive quiescent galaxy at these redshifts so far, it is not possible to determine how typical these properties are of the broader population.

One thing that does appear to be typical of the $z>3$ massive quiescent galaxy population is extremely small physical sizes (e.g., \citealt{Ito2023, Wright2023, Ji2024}). This represents an extension of the well-known trend at $z < 3$ towards smaller sizes for quiescent galaxies with increasing redshift (e.g., \citealt{Daddi2005, Trujillo2006, McLure2013, vanderwel2014, Mowla2019, Hamadouche2022}), and is also consistent with the finding that the youngest quiescent galaxies at cosmic noon are the smallest (e.g., \citealt{Whitaker2012, Almaini2017}). The most extreme example known is again GS-9209, with an effective radius, $r_e \simeq 200$ pc and a stellar mass surface density within $r_e$ of log$_{10}(\Sigma_\mathrm{eff}/$M$_\odot$ kpc$^{-2}) = 11.15\pm0.08$. It seems likely that these extremely small sizes must be related to the ability of these galaxies to form their large stellar masses in such a short space of time (e.g., \citealt{Dekel2023}), as well as the subsequent shutting down of star formation.

Recently, \cite{Glazebrook2023} have reported a quiescent galaxy, ZF-UDS-7329 at $z\simeq3.2$, for which they derive a stellar mass of log$_{10}(M_*/\mathrm{M_\odot})=11.26^{+0.03}_{-0.16}$. They further report that the bulk of this mass formed at $z\simeq11$, approximately 400 Myr after the Big Bang. The authors claim that, given the area of imaging data from which this galaxy was selected, the presence of such a large stellar mass within a single dark matter halo at $z\simeq11$ would be strongly in tension with $\Lambda$-CDM cosmology (see also \citealt{Glazebrook2017, Antwi-Danso2023, deGraaff2024,UrbanoStawinski2024}).

There are several potential means of resolving this tension, in particular, a.) an exotic, top-heavy stellar initial mass function (IMF), meaning that the stellar mass is in fact lower; b.) very high average stellar metallicity, meaning the stellar population is in fact younger; or c.) this galaxy having formed via hierarchical mergers between galaxies that each individually formed in separate halos around $z\simeq11$. \cite{Glazebrook2023} argue that all three of these scenarios are unlikely, and suggest that revisions may be necessary to our current understanding of dark matter and hence structure formation in the early Universe.

It seems clear that two key questions must now be answered. Firstly, what causes so many massive galaxies to experience a rapid shutting down of star-formation activity at these early epochs? Secondly, does the number density of massive (and potentially already old) quiescent galaxies at $z > 3$ require revisions to current models of galaxy formation physics, or even $\Lambda$-CDM cosmology? 

Answering these questions is one of the key motivations for the JWST Early eXtragalactic Continuum and Emission Line Science (EXCELS) NIRSpec Cycle 2 survey (Programme ID: 3543; PI: Carnall; Co-PI: Cullen). This programme was designed as a direct follow up to both the JWST Public Release IMaging for Extragalactic Research (PRIMER) Cycle 1 programme, as well as our Cycle 1 NIRSpec observations of GS-9209.

The primary aim of EXCELS is to expand high-SNR, medium-resolution continuum spectroscopy to a representative sample of massive quiescent galaxies at $3 < z < 5$. Because these objects are relatively rare, we are also able to dedicate substantial space on the NIRSpec Micro-Shutter Array (MSA; \citealt{Ferruit2022}) to observing a variety of other key target classes, in particular high-redshift star-forming galaxies, as well as quiescent galaxies at the cosmic noon epoch ($1 < z < 3$).

In this work, we introduce the EXCELS survey and present the first scientific results. We focus on the most massive quiescent galaxies at $3 < z < 5$, in particular the 4 such objects in EXCELS that have log$_{10}(M_*/\mathrm{M_\odot}) > 11$. Objects in this mass range are often known as ultra-massive galaxies (e.g., \citealt{Forrest2020}). We focus on these most-massive objects in particular as these are the most likely to present a problem for galaxy formation models and/or $\Lambda$-CDM, allowing us to critically examine the debate in the literature on this topic to date.

This sample of 4 objects includes a pair of galaxies at $z=4.62$ physically separated by $860$ pkpc, an extreme post-starburst galaxy (PSB) at $z=3.99$, and the $z=3.19$ object ZF-UDS-7329 discussed by \cite{Glazebrook2023}. We infer the physical properties of these galaxies via full spectral fitting, in particular their SFHs, and consider whether or not these are consistent with the $\Lambda$-CDM halo-mass function using the extreme value statistics (EVS) framework of \cite{Lovell2023}.

In Section \ref{sec:excels_overview}, we provide a high-level overview of the EXCELS survey design and observing strategy. In Section \ref{sec:excels_sample} we provide a detailed description of the EXCELS target-selection and mask-design procedures, including the selection of high-redshift massive quiescent galaxies from the JWST PRIMER imaging survey. Our data reduction and fitting methodologies are described in Section \ref{sec:methods}. We then present our results in Section \ref{sec:results}, discuss these results in Section \ref{sec:discussion}, and present our conclusions in Section \ref{sec:conclusion}.

All magnitudes are quoted in the AB system. For cosmological calculations, we adopt $\Omega_M = 0.3$, $\Omega_\Lambda = 0.7$ and $H_0$ = 70 $\mathrm{km\ s^{-1}\ Mpc^{-1}}$. We assume a \cite{Kroupa2001} initial mass function, and assume the Solar abundances of \cite{Asplund2009}, such that $\mathrm{Z_\odot} = 0.0142$.

\section{EXCELS Observing Strategy}\label{sec:excels_overview}

EXCELS fundamentally consists of four NIRSpec MSA pointings within the PRIMER Ultra-Deep Survey (UDS) field. Each of these is observed with three medium-resolution grating/filter combinations: G140M/F100LP, G235M/F170LP and G395M/F290LP, providing continuous wavelength coverage from $1-5$ $\mathrm{\mu m}$ at spectral resolving power, $R=\lambda/\Delta\lambda \simeq1000$. 

Separate MSA configurations are specified for each of the three gratings within each pointing. This strategy maximises the number of objects for which we can observe the key rest-frame near-UV to optical wavelength range of interest ($\lambda=3600-7000$\AA). For example, a $z\simeq1$ quiescent galaxy can occupy the same space on the MSA when observing with G140M that is occupied by a $z>3.5$ star-forming galaxy when observing with G235M and G395M.

Observations were conducted using a 3-point nodding pattern within 3-shutter MSA slitlets. Additional shutters were opened manually where possible to extend these slitlets, and additional sky slitlets were opened in unused areas of the MSA to provide the option of performing a master background subtraction. The total integration times for each grating within each pointing are 14706 seconds ($\simeq4$ hours) in G140M and G395M, and 19958 seconds ($\simeq5.5$ hours) in G235M, using the NRSIRS2 readout pattern.

\section{EXCELS Sample Selection}\label{sec:excels_sample}

The EXCELS sample selection process is relatively involved, drawing together several different sub-samples of target objects, with different prioritisation weighting schemes for each of the three gratings.

At its most basic level, the EXCELS sample is drawn from two photometric catalogues. Firstly, the VANDELS survey UDS-HST selection catalogue \citep{McLure2018b}, which is heavily based on the CANDELS UDS catalogue of \cite{Galametz2013}. Secondly, a new catalogue based on the JWST Cycle 1 PRIMER UDS imaging (\citealt{Dunlop2021}; McLeod et al. in prep). The four key target classes within these catalogues that the EXCELS survey is built around are:

\begin{itemize}\setlength\itemsep{6pt}
\item PRIMER massive quiescent galaxies at $2 \leq z \leq 5$
\item VANDELS massive quiescent galaxies at $1 \leq z \leq 2.5$
\item VANDELS star-forming galaxies at $2.4 \leq z \leq 7$
\item PRIMER star-forming galaxies at $z \geq 5$.
\end{itemize}

\noindent We describe these two catalogues in more detail in Sections \ref{subsec:vandels} and \ref{subsec:primer}, as well as the processes by which the four EXCELS target galaxy samples were selected from them. We then describe the process by which these samples were placed onto the NIRSpec MSA in Section \ref{subsec:msa_config}.

\subsection{VANDELS UDS-HST catalogue and selection}\label{subsec:vandels}

VANDELS \citep{McLure2018b, Pentericci2018, Garilli2021} is a large European Southern Observatory (ESO) public spectroscopic survey on the Very Large Telescope (VLT) Visible Multi-Object Spectrograph (VIMOS; \citealt{LeFevre2003}). The survey focused on the high-redshift Universe, with the vast majority ($\simeq97$ per cent) of targets being drawn from the following three categories:

\begin{itemize}\setlength\itemsep{6pt}
\item Bright star-forming galaxies at $2.4 \leq z \leq 5.5$
\item Lyman break galaxies at $3 \leq z \leq 7$
\item Massive quiescent galaxies at $1 \leq z \leq 2.5$.
\end{itemize}

\noindent The survey targeted both the UDS and GOODS-South fields, building upon the legacy HST imaging in these fields from CANDELS \citep{Grogin2011, Koekemoer2011}. The parent photometric sample was drawn from four photometric catalogues, described in full detail in \cite{McLure2018b}.

For EXCELS, we consider only the VANDELS UDS-HST catalogue, based on the \cite{Galametz2013} UDS CANDELS catalogue. We include all objects photometrically selected as potential VANDELS spectroscopic targets in the above three categories. However, we give priority to those that were eventually observed as part of VANDELS (see Section \ref{subsec:msa_config}). The full definitions of these three samples are given in \cite{McLure2018b}; we here provide a brief summary.

\subsubsection{VANDELS star-forming galaxies at $2.4 \leq z \leq 7$}

The sample of VANDELS star-forming galaxies we include in EXCELS is a combination of the VANDELS bright star-forming and Lyman break galaxy (LBG) samples, which we treat equally in our prioritisation scheme. These VANDELS star-forming samples were selected firstly by the two photometric redshift criteria given above. In the UDS-HST region targeted by EXCELS, the photometric redshifts used for all object classes were those produced by the CANDELS team by taking the median of results obtained with a variety of different photometric redshift codes \citep{Dahlen2013}.

The bright star-forming sample was then selected by $i-$band magnitude, requiring $i \leq 25$, in order to ensure high SNR in the VANDELS VIMOS spectra. The LBG sample comprises fainter star-forming galaxies: in the UDS-HST region, LBGs at $3 \leq z \leq 5.5$ were required to have $25 \leq H \leq 27$ and $i \leq 27.5$. In a slight variation, LBGs at $5.5 \leq z \leq 7$ were instead required to have $25 \leq H \leq 27$ and $z^\prime \leq 26.5$, due to the impact of intergalactic medium (IGM) attenuation on the $i$-band photometry for these objects.

\subsubsection{VANDELS massive quiescent galaxies at $1 \leq z \leq 2.5$}

The VANDELS quiescent sample at $1 \leq z \leq 2.5$ was selected firstly by CANDELS photometric redshift, then by requiring $H \leq 22.5$ and $i\leq25$. This $H$-band magnitude limit roughly corresponds to a stellar-mass limit of log$_{10}(M_*/\mathrm{M_\odot})~=~10$ at $z\lesssim1.5$ where the majority of the sample is located. As with the star-forming sample, the $i-$band limit again guarantees high SNR in the VANDELS VIMOS spectra. Quiescent objects were then selected by rest-frame UVJ colour, using the $z>1$ criteria of \cite{Williams2009}.

\begin{figure*}
	\includegraphics[width=0.98\textwidth]{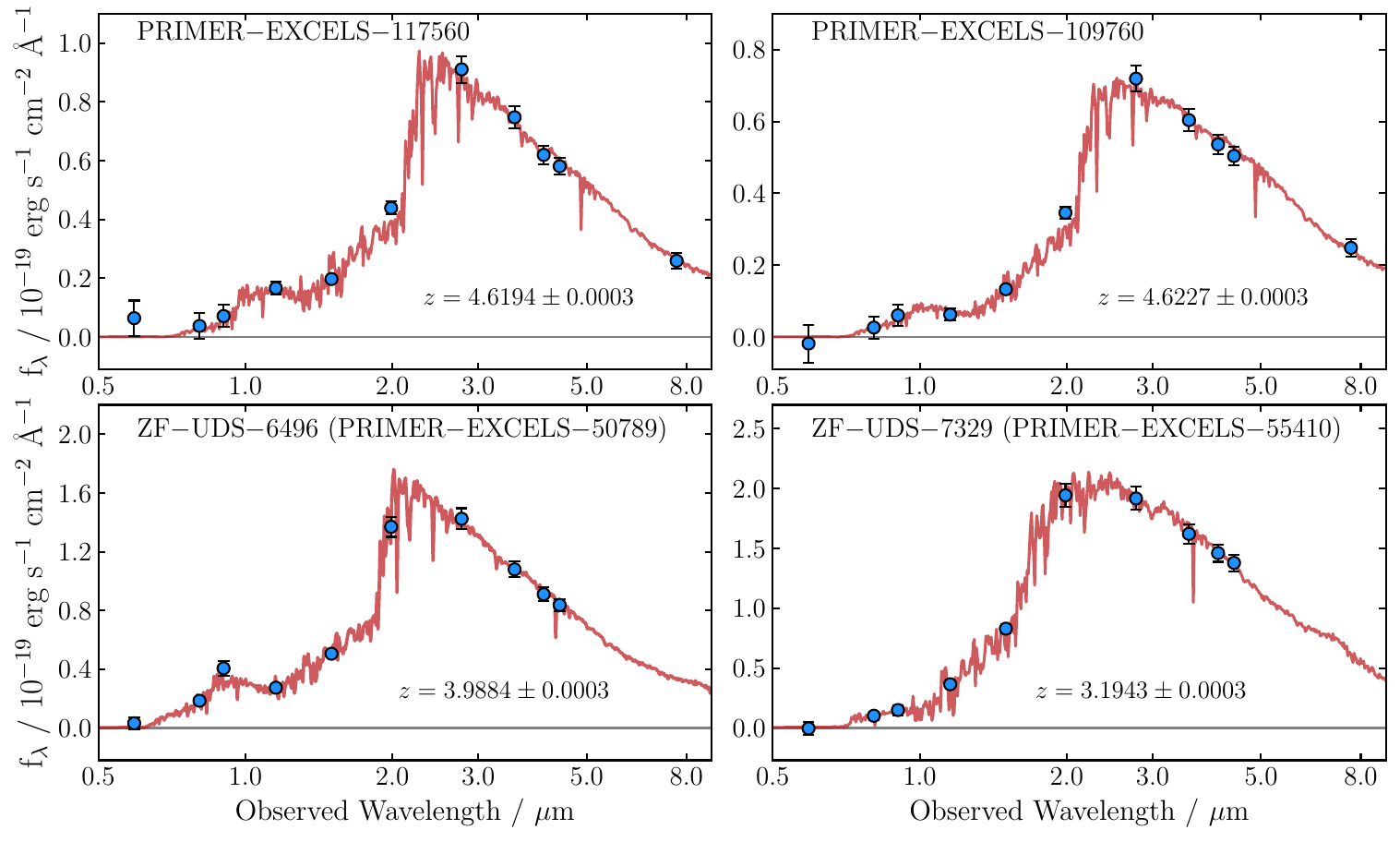}
    \caption{Spectral energy distributions (SEDs) for the 4 most massive $z>3$ quiescent galaxy candidates selected from our PRIMER catalogue by the process described in Section \ref{subsec:primer_passive}. These are the only 4 candidates at $z>3$ that have inferred stellar masses, log$_{10}(M_*/\mathrm{M_\odot}) > 11$. Photometric data  are shown with blue points and our posterior-median fitted models with red lines. All 4 were observed spectroscopically as part of EXCELS, and their spectra are shown in Fig. \ref{fig:spec2}. The top two are our newly confirmed $z=4.62$ galaxies. The bottom two were previously selected by \protect\cite{Schreiber2018} and spectroscopically confirmed by \protect\cite{Nanayakkara2024}. The top two objects also have PRIMER MIRI coverage, and we include the F770W fluxes, which are consistent with our fitted models.}
    \label{fig:spec1}
\end{figure*}

\subsection{PRIMER UDS catalogue and selection}\label{subsec:primer}

PRIMER \citep{Dunlop2021} is a large public JWST NIRCam+MIRI Cycle 1 treasury imaging programme. The survey targets the UDS and COSMOS fields, providing contiguous NIRCam coverage totalling $\simeq400$ sq. arcmin across both fields. Much of this area overlaps with existing HST imaging. The NIRCam imaging spans a wavelength range from $1-5$ $\mathrm{\mu m}$ in 8 photometric bands: F090W, F115W, F150W, F200W, F277W, F356W, F410M and F444W. PRIMER also provides MIRI imaging in the F770W and F1800W bands over approximately half of the NIRCam area.

To select galaxies for EXCELS, a PRIMER UDS NIRCam catalogue was produced. The imaging was reduced using the PRIMER Enhanced NIRCam Image Processing Library (PENCIL; Magee et al. in prep) v0.6, a custom version of the JWST pipeline (v1.10.2). We made use of the CRDS\_CTX = jwst\_1118.pmap version of the JWST calibration files. We also produced custom mosaics in the HST ACS F435W, F606W and F814W bands using data from CANDELS \citep{Grogin2011, Koekemoer2011}. All bands were point spread function (PSF) homogenised to the F444W band using kernels based on stacking bright stars in the field. 

A catalogue was then produced using the \textsc{Source Extractor} software \citep{Bertin1996} in dual image mode, using F356W as the selection band. The long-wavelength F356W band was chosen in order to select objects as much as possible by stellar mass, rather than ongoing star formation, which dominates at shorter wavelengths. The F444W band was not used as it is substantially shallower. Fluxes were extracted in small, 0.35$^{\prime\prime}$-diameter apertures optimised for the extremely compact high-redshift galaxies we intended to select. We then scaled these up to total fluxes using the F356W FLUX\_AUTO values measured by \textsc{Source Extractor}. We included an error floor in each band for each object at 5 per cent of the observed flux, to account for potential flux calibration systematic uncertainties (e.g., \citealt{Brammer2008}).

The final catalogue covers an area of $\simeq 160$ sq. arcmin and is restricted to those 42905 objects with coverage in all 11 HST+NIRCam photometric bands. Following the selection of the final 4 massive quiescent galaxies at $3<z<5$ that are the focus of this work, additional MIRI F770W photometry was also extracted for the galaxy pair at $z=4.62$. The other two galaxies lie outside the MIRI footprint. This photometry was initially measured in 0.5$^{\prime\prime}$-diameter apertures and then aperture corrected using a kernel that matched the F444W and F770W PSFs.

\subsubsection{PRIMER massive quiescent galaxies at $2 < z < 5$}\label{subsec:primer_passive}

To select high-redshift massive quiescent galaxies from our combined 11-band HST ACS + JWST NIRCam catalogue, we follow a probabilistic selection method almost identical to that presented in \cite{Carnall2020, Carnall2023b}. This method has recently been validated via a series of different spectroscopic follow-up campaigns \citep{Carnall2023c,deGraaff2024,Barrufet2024}.

We begin by fitting all galaxies in the catalogue with F356W $\leq 25$ using the \textsc{Bagpipes} code \citep{Carnall2018, Carnall2019b}. This limit was chosen to ensure sufficient SNR in the EXCELS spectra to extract stellar ages and metallicities via full spectral fitting (typically assumed to be $\gtrsim30$ per resolution element at $R\simeq1000$; e.g., \citealt{Pacifici2012}).

We employ the same model configuration as used in the above papers. Briefly, this consists of the 2016 updated version of the \cite{Bruzual2003} stellar population models using the MILES stellar spectral library \citep{Sanchez-Blazquez2006, Falcon-Barroso2011} and a double-power-law SFH model (e.g., \citealt{Carnall2019a}). The effects of this choice are discussed in Section \ref{subsec:disc_sfhs} and Appendix \ref{app1}. Emission lines are included using the \textsc{Cloudy} photoionization code \citep{Ferland2017} with an approach based on that of \cite{Byler2017}. Dust attenuation is included using the variable-slope model of \cite{Salim2018}, which is based on the \cite{Calzetti2000} curve. IGM attenuation is included using the model of \cite{Inoue2014}. The model posterior is sampled using the \textsc{MultiNest} \citep{Feroz2019} code, accessed via the \textsc{PyMultiNest} package interface \citep{Buchner2014}.

To select quiescent objects, we apply a sSFR criterion, requiring sSFR $< 0.2$ / $t_\mathrm{H}(z)$, where $t_\mathrm{H}(z)$ is the age of the Universe as a function of observed redshift. This criterion has been widely applied in the literature (e.g., \citealt{Gallazzi2014, Pacifici2016}), and has been shown to be virtually equivalent to a rest-frame UVJ colour cut (e.g., \citealt{Carnall2018, Carnall2020, Fang2018}).

In a slight refinement of our previous work, we select quiescent galaxies based on the 2D joint posterior distribution of redshift and specific star-formation rate (sSFR), rather than imposing separate criteria on the 1D posteriors for each of these parameters. For each sample in the posterior for each object, we calculate whether the sSFR is below the threshold quoted above. We then require that 50 per cent of posterior samples are both above $z=2$ and below our sSFR threshold.

We then further define a robust sub-sample of these galaxies by requiring that 95 per cent of the joint posterior samples are both at $z>1.75$ and below the sSFR threshold. We use $z>1.75$  rather than $z>2$ when selecting our robust sub-sample to make sure that objects at $z\simeq2$ with redshift posteriors that extend slightly below $z=2$ are not excluded.

We finally visually inspect the whole sample in order to exclude any imaging artefacts and contaminants. This process results in a sample of 127 candidates (78 robust), of which 100 (65 robust) are at $2 < z < 3$, a further 21 (10 robust) are at $3 < z < 4$, and the final 6 (3 robust) are at $4 < z < 5$. We do not identify any candidates at $z>5$, though we do not impose any upper limit on redshift for our sample. 

In Fig. \ref{fig:spec1} we show spectral energy distributions (SEDs) for the 4 most massive candidates at $z>3$ identified by this process. We present a detailed analysis of the EXCELS spectra for these galaxies in Section \ref{sec:results}.

\subsubsection{PRIMER star-forming galaxies at $z \geq 5$}\label{subsec:primer_hiz}

To select our high-redshift star-forming sample we begin with the same PRIMER UDS catalogue described in Section \ref{subsec:primer_passive}, and follow a procedure very similar to those laid out in \cite{Donnan2022} and \cite{McLeod2024}. We here provide a brief summary. The initial sample consists of all galaxies with $z_\mathrm{med}\geq5$, where $z_\mathrm{med}$ is the median photometric redshift from 5 runs with different public codes and configurations. Two of these runs used the \textsc{LePhare} code \citep{Arnouts1999, Ilbert2006} with the \cite{Bruzual2003} and PEGASE.2 \citep{Fioc1999} templates. Two further runs were conducted with \textsc{EAZY} \citep{Brammer2008} using the PCA and PEGASE.2 templates. A final run was conducted using the photometric redshift code described in \cite{McLure2011}.

These galaxies are then re-analysed using an updated version of the code described in \cite{McLure2011}. The updates are designed to reproduce the extreme emission-line fluxes observed in high-redshift star-forming galaxies. This re-analysis is performed to improve the photometric-redshift accuracy of high equivalent width (EW) line emitters, with a particular focus on objects where strong line emission contaminates the F410M band (e.g., [O\,\textsc{iii}] 5008\AA\ at $6.7<z<7.6$). Objects of this nature can return erroneous photometric redshifts if the high EW line emission is not properly accounted for. 

Following this re-analysis, all candidates with robust photometric redshifts in the range $5\leq z \leq 8$ are regarded as potential spectroscopic targets, where robust means $\Delta \chi^2\geq 4$ between the primary redshift solution and any alternative solutions at lower redshift. Finally, all candidates are visually inspected to remove any remaining artefacts or contaminants.

We supplement the sample at $z\geq8$ using an alternative PRIMER UDS catalogue described in \cite{Donnan2024}. The full sample selection process is described in that work, we here provide a brief summary. This catalogue uses smaller, $0.3^{\prime\prime}$-diameter apertures, and uses the F150W, F200W and F277W filters as detection images. This configuration is better optimised to the selection of galaxies at $z\geq8$. We begin by selecting objects with a non-detection above a $2\sigma$ threshold in the HST/ACS filters as well as the JWST/NIRCam F090W filter, as these are all blueward of the Lyman break at $z>8$.

We then require detections of flux redward of the Lyman break, at a $\geq8\sigma$ level in the first filter and a $\geq5\sigma$ level in the next filter. The photometry of the candidates was then fitted using the \textsc{Eazy} code \citep{Brammer2008} with the \textsc{Pegase} \citep{Fioc1999} set of templates. We then require a best-fitting photometric redshift of $z\geq8$ and a $\Delta \chi^2\geq4$ between the best-fitting high-redshift and next-most-probable lower-redshift solutions. All candidates are finally visually inspected to remove artefacts and contaminants.

\subsection{Pointing locations and MSA configuration}\label{subsec:msa_config}

The source prioritisation scheme used to define our MSA configurations includes 10 separate priority classes within our 4 target samples (described at the start of Section \ref{sec:excels_sample}), with each class being assigned half the weight of the class above it. We split both PRIMER samples into different priority classes depending on redshift. We split both VANDELS selection catalogue samples based on redshift and whether the galaxy was actually observed as part of VANDELS. The VANDELS star-forming sample was then further split based on the spectroscopic redshift quality flag, zflag, assigned by the VANDELS team (using the scheme of \citealt{LeFevre2005}; see \citealt{Pentericci2018}). Our 10 priority classes are:

\begin{enumerate}\setlength\itemsep{6pt}
\item PRIMER quiescent ($z > 4$)
\item PRIMER quiescent ($3.5 < z < 4$)
\item PRIMER quiescent ($3 < z < 3.5$)
\item VANDELS obs star-forming ($z>3$ \& $2\leq$ zflag $\leq9$)
\item VANDELS obs quiescent ($1 < z < 1.3$)
\item PRIMER quiescent ($2 < z < 3$)
\item PRIMER star-forming ($z > 6.7$)
\item VANDELS selection quiescent (all except class v)
\item PRIMER star-forming ($5 < z < 6.7$)
\item VANDELS selection star-forming (all except class iv)
\end{enumerate}

\noindent Where there is overlap between these samples (e.g., a $z=5.25$ star-forming galaxy appearing in both the VANDELS and PRIMER samples), objects were prioritised based on the highest-priority category in which they appear.

We selected the locations of our four EXCELS MSA pointings via a three-step process. Firstly, rough locations were chosen by eye from a plot of the PRIMER UDS field, showing the locations of the highest-priority targets. We then employed the eMPT software \citep{Bonaventura2023} to perform a search over a relatively wide area (30$^{\prime\prime}\times30^{\prime\prime})$ for the best pointing locations. We finally loaded our catalogue into the Astronomer's Proposal Tool (APT) software and used the MSA Planning Tool (MPT) to perform a much finer search (a 1$^{\prime\prime}\times1^{\prime\prime}$ grid with points at $0.01^{\prime\prime}$ intervals) centred on the eMPT best pointing coordinates. Using this process, we found that it was usually possible to match, or even slightly better, the eMPT solution in MPT.

The first two of the above steps were performed once for each pointing, including all 10 of the above priority classes. The third step in APT was however performed separately for all three gratings within each pointing, including only the subset of the above priority classes that we wished to observe in that grating. This strategy ensures that we observe as many high-priority targets as possible in the relevant gratings, whilst retaining the spatial overlap required to observe individual objects in multiple gratings where necessary.

This scheme was arrived at after extensive simulation, with the aim of maximising the number of scientifically interesting targets whilst retaining a broad spread between our different target classes. In general, we prioritise higher-redshift galaxies over lower-redshift ones within our four target classes. The split in redshift for the VANDELS quiescent sample prioritises objects at $1 < z < 1.3$, as the mass-completeness limit is lowest (log$_{10}(M_*/\mathrm{M_\odot}) = 10.3$, see \citealt{Carnall2019b}) at the low end of the VANDELS redshift range, providing a representative sample over a large dynamic range in mass. We prioritise galaxies from the VANDELS selection catalogue that have VANDELS spectra to ensure the largest possible samples for joint analyses. We prioritise VANDELS star-forming galaxies with $2 \leq$ zflag $\leq9$ as these have the highest SNRs in their VANDELS spectra.

Not all of our 10 priority classes are included when designing the MSA configurations for all three gratings. This is because only $1-2$ gratings are required to cover the key rest-frame near-UV to optical wavelength range of interest ($3600-7000$\AA) for an individual galaxy. For the G140M grating, all priority classes except class ix are included, with galaxies of class iv only included if they have VANDELS spectroscopic redshifts, $z_\mathrm{spec} < 3.6$. Priority class vii was included in our G140M mask design in order to provide coverage of the Lyman$-\alpha$ line for these highest-redshift candidates, whereas for galaxies of class ix at $5 < z < 6.7$ we expect that Lyman$-\alpha$ will not fall within the G140M wavelength range. For the G235M grating, priority classes i-iv, vi, ix and x are included. For the G395M grating, priority classes i-iv, vii, ix and x are included, with galaxies of class iv only included if they have $z_\mathrm{spec} > 3.4$. 

We finally include a separate filler list for each grating from the PRIMER catalogue, below our 10 priority classes. For G140M, we include objects with F356W $< 26$ and photometric redshifts $0.5 < z_\mathrm{phot} < 2.5$ (using the median photometric redshifts described in Section \ref{subsec:primer_hiz}). For G235M, we include objects with F356W $< 26$ and $1.5 < z_\mathrm{phot} < 5$. For G395M, we include objects with F356W $< 28$ and $z_\mathrm{phot} > 3.5$, whilst giving higher priority to objects with F356W $< 26$. These redshift intervals were chosen to target bright rest-frame optical emission lines for filler targets.

\begin{figure*}
	\includegraphics[width=0.95\textwidth]{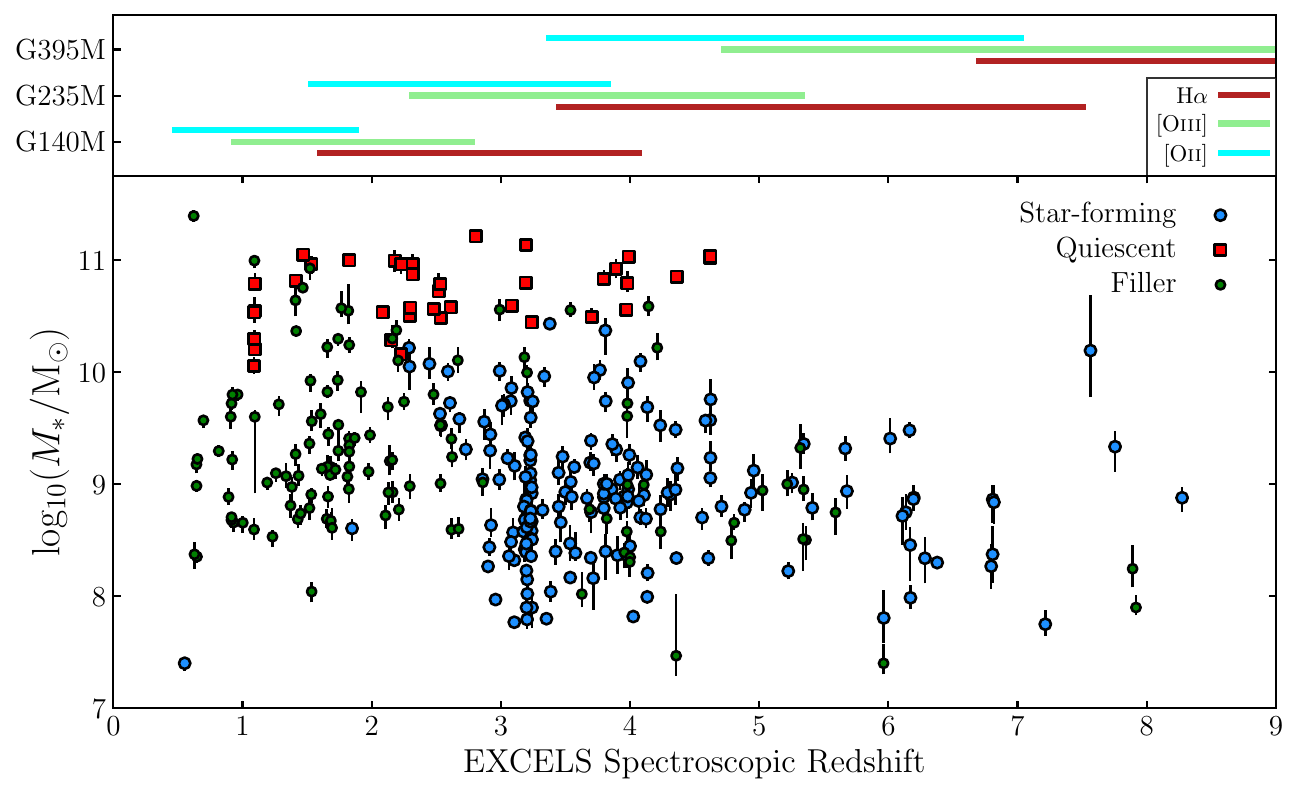}
    \caption{Stellar mass vs redshift distribution for the 341 out of 401 EXCELS galaxies that could be assigned secure (flag 3, 4 or 9) spectroscopic redshifts (see Section \ref{subsec:method_specz}). Stellar masses were measured by repeating the photometric fitting analysis described in Section \ref{subsec:primer_passive} with redshifts fixed to the values derived from the EXCELS spectra. The EXCELS sample has been split into quiescent galaxies (red squares), star-forming galaxies (blue large circles) and filler galaxies (green small circles). The selection of these sub-samples is described in Section~\ref{sec:excels_sample}. The bars at the top show the redshift ranges over which key spectral features are visible in the three medium-resolution gratings used. A full list of EXCELS galaxy IDs, coordinates, spectroscopic redshifts and their associated quality flags is given in Table \ref{table:redshifts}.}
    \label{fig:z_vs_mass}
\end{figure*}

\section{Methods}\label{sec:methods}

\subsection{Data Reduction}

We reduce the EXCELS spectra for use in this paper as follows. We begin by downloading the raw level 1 ``uncal'' products from the Mikulski Archive for Space Telescopes (MAST). We process these files using v1.12.5 of the JWST data reduction pipeline\footnote{https://github.com/spacetelescope/jwst}, and make use of the CRDS\_CTX = jwst\_1183.pmap version of the JWST Calibration Reference Data System (CRDS) files. We run the level 1 pipeline using the default configuration, except for turning on the advanced snowball rejection option (by setting the expand\_large\_events flag to True). 

We run the level 2 pipeline using the default configuration options, except for turning off the sky subtraction step. We then manually flag and mask any remaining artefacts in the separate 2D calibrated spectra for each of our three exposures taken at different nod positions. We then perform the sky subtraction step on the masked 2D spectra for each exposure using custom code. This step produces the inputs required for the level 3 pipeline, which we then run using the default configuration options to produce final combined 2D flux calibrated spectra.

We then perform our own custom 1D optimal extraction of the 2D spectra \citep{Horne1986}. We set the extraction centroid by calculating the flux-weighted mean position of the object, considering only the region that falls within the central NIRSpec MSA slitlet in the PRIMER F356W imaging. We make use of the \textsc{MsaExp} code\footnote{https://github.com/gbrammer/msaexp} to plot the spatial extent of the slitlets on top of the PRIMER imaging. We calculate the optimal extraction weights by performing Gaussian fits to our wavelength-collapsed 2D spectra, allowing the centroid measured from the PRIMER imaging to deviate by up to 2 pixels in either direction (though the typical adjustment is much smaller than this).

This is all of the data processing necessary to begin our spectroscopic redshift measurement process, which is described in Section \ref{subsec:method_specz}. For the 4 ultra-massive quiescent galaxies discussed in Section \ref{sec:results}, we then apply the final processing steps below to facilitate our full spectral fitting analysis, which is described in Section \ref{subsec:spec_fitting}.

We combine our three separate gratings by first calculating the mean flux in the overlapping wavelength regions, then scaling the G140M and G395M spectra to the G235M normalisation. We degrade the resolution of the higher-resolution (shorter-wavelength) grating for each of the two overlap regions to that of the lower-resolution (longer-wavelength) grating using \textsc{SpectRes} \citep{Carnall2017}. We then combine the data in the overlap regions by averaging pixel values between the two gratings.

At this stage we do not perform any modification to account for slit losses or imperfect spectrophotometric calibration (the JWST absolute flux calibration process applied in the pipeline, which includes a point-source slit-loss calculation, is described in \citealt{Gordon2022}). Instead we fit for these effects by introducing nuisance parameters into our full-spectral-fitting analysis, as described in Section \ref{subsec:spec_fitting}.

\begin{table}

\caption{Spectroscopic redshifts for the EXCELS galaxies, measured as described in Section \ref{subsec:method_specz}. Redshift quality flags have been assigned following the scheme set out in \protect \cite{LeFevre2005}. The full table is available as supplementary online material.}
\begingroup
\setlength{\tabcolsep}{6pt}
\renewcommand{\arraystretch}{1.1}
\begin{center}

\begin{tabular}{lcccc}
\hline
ID & RA / deg & DEC / deg & $z_\mathrm{spec}$ & Quality flag  \\
\hline
34495 & 34.289463 & $-$5.269810 & 3.797 & 4 \\
35221 & 34.301028 & $-$5.268660 & 1.742 & 4 \\
37445 & 34.278294 & $-$5.264346 & 2.447 & 4 \\
39063 & 34.290443 & $-$5.262081 & 3.703 & 4 \\
39138 & 34.276844 & $-$5.259679 & 1.737 & 4 \\
& & ... & & \\
\hline
\end{tabular}
\end{center}
\endgroup
\label{table:redshifts}
\end{table}

\subsection{Spectroscopic redshift and stellar mass measurements from EXCELS}\label{subsec:method_specz}

We measure spectroscopic redshifts from the EXCELS data manually, using the \textsc{Pandora} software for data visualisation \citep{Garilli2010}. Two sets of separate measurements were made by different team members, and the results reconciled, with quality flags assigned following the framework described in \cite{LeFevre2005,Pentericci2018}. A high-quality redshift flag of 3, 4 or 9 could be assigned for 341 of the 401 objects observed, with 309 objects assigned a flag 4 redshift, corresponding to $\gtrsim99$ per cent confidence. A list of the derived redshifts for the 349 EXCELS objects with a spectroscopic redshift of any quality are given in Table \ref{table:redshifts}.

Following spectroscopic redshift measurement, we re-run the photometric fitting process described in Section \ref{subsec:primer_passive} for all 341 objects with high-quality redshift measurements, whilst fixing their redshifts to the spectroscopic values. The stellar mass vs redshift distribution for the EXCELS sample derived in this way is shown in Fig. \ref{fig:z_vs_mass}.

\begin{table*}

\caption{The 11 free parameters of the \textsc{Bagpipes} model we fit to our spectroscopic+photometric data for our four ultra-massive quiescent galaxies, along with their associated prior distributions. The model is described in Section \ref{subsec:spec_fitting} and the fits to the data are shown in Fig~\ref{fig:spec2}.  The upper limit on $\tau$, $t_\mathrm{obs}$, is the age of the Universe as a function of redshift. Logarithmic priors are all applied in base ten.}
\begingroup
\setlength{\tabcolsep}{6pt}
\renewcommand{\arraystretch}{1.1}
\begin{tabular}{llllllll}
\hline
Component & Parameter & Symbol / Unit & Range & Prior & \multicolumn{2}{l}{Hyper-parameters} \\
\hline
General & Redshift & $z$ & ($z_\mathrm{spec}-0.05$, $z_\mathrm{spec}+0.05$) & Gaussian & $\mu=z_\mathrm{spec}$ & $\sigma=0.01$ \\
& Stellar velocity dispersion & $\sigma$ / km s$^{-1}$ & (50, 500) & Logarithmic & & \\
\hline
SFH & Total stellar mass formed & $M_*\ /\ \mathrm{M_\odot}$ & (1, $10^{13}$) & Logarithmic & & \\
& Stellar metallicity & $Z_*\ /\ \mathrm{Z_\odot}$ & (0.00355, 3.55) & Logarithmic & & \\
& Double-power-law falling slope & $\alpha$ & (0.1, 1000) & Logarithmic & & \\
& Double-power-law rising slope & $\beta$ & (0.1, 1000) & Logarithmic & & \\
& Double-power-law turnover time & $\tau$ / Gyr & (0.1, $t_\mathrm{obs}$) & Uniform & & \\
\hline
Dust & $V-$band attenuation & $A_V$ / mag & (0, 4) & Uniform & & \\
& Deviation from Calzetti slope & $\delta$ & ($-0.3$, 0.3) & Gaussian & $\mu = 0$ & $\sigma$ = 0.1 \\
& Strength of 2175\AA\ bump & $B$ & (0, 5) & Uniform & & \\

\hline

Noise & White noise scaling & $a$ & (0.1, 10)  & logarithmic & & \\
\hline
\end{tabular}
\endgroup
\label{table:params}
\end{table*}

\subsection{Spectrophotometric fitting}

\subsubsection{Bagpipes fitting}\label{subsec:spec_fitting}

For our 4 EXCELS ultra-massive quiescent galaxies at $3 < z < 5$ we perform full spectral fitting on our spectroscopic data, in combination with the available HST+JWST photometry, again using the \textsc{Bagpipes} code. The model configuration includes all the same components described in Section \ref{subsec:primer_passive} for fitting our photometric catalogue, with some additions that are necessary due to the inclusion of spectroscopic data. A full list of fitted model parameters and their associated priors is given in Table \ref{table:params}.

We first restrict our spectroscopic data to rest-frame wavelengths $3540-7350$\AA\ for the purpose of full spectral fitting. This is the wavelength range spanned by the MILES library in the \cite{Bruzual2003} models, with the models available outside this range being of lower spectral resolution. This wavelength range contains all of the key age-sensitive features necessary to constrain the SFH.

We include a multiplicative high-order Chebyshev polynomial in our model for our spectroscopic data, to account for both slit losses and any imperfections in spectrophotometric calibration (e.g., \citealt{Cappellari2017}). This polynomial is optimised analytically at each step in the posterior sampling process (e.g., \citealt{Johnson2021}). We include one polynomial order per 100\AA\ of rest-frame wavelength coverage. This is often taken as a rule of thumb, as it allows reasonable flexibility to calibrate the spectroscopic data to the available photometry without allowing the polynomial to reproduce individual emission and absorption features in galaxy spectra (e.g., \citealt{Conroy2012, Conroy2014}).

This results in a 38\textsuperscript{th} order polynomial for these data. However we have also experimented with higher (76\textsuperscript{th}) and lower (10\textsuperscript{th}, 19\textsuperscript{th}) orders, as well as a low (2\textsuperscript{nd}) order polynomial with full Bayesian optimisation of each coefficient (as used in our previous work; e.g., \citealt{Carnall2023c}) to verify that this choice does not significantly affect our results. As we have access to high-SNR PRIMER NIRCam photometry across our full spectroscopic wavelength range (see Fig. \ref{fig:spec1}), the calibration polynomial is very well constrained. We typically make only minor relative modifications to the calibration of the spectra as a function of wavelength, typically at the $\pm5$ per cent level. We observe that, following the application of our fitted polynomial, the integrated fluxes across the relevant wavelength ranges in our spectra are well-matched with the corresponding photometry.

We also include a Gaussian convolution of our model spectrum to account for the effects of velocity dispersion in our target galaxies. We assign this a width, $\sigma$, which we allow to vary between $50-500$ km s$^{-1}$ with a logarithmic prior. We finally include a multiplicative factor on the error spectrum for our spectroscopic data to account for potentially underestimated uncertainties (e.g., see \citealt{Maseda2023})\footnote{see also https://github.com/spacetelescope/jwst/issues/7362}, which we allow to vary from $a=1-10$ with a logarithmic prior. 

As discussed in Section \ref{sec:results}, there is substantial evidence that the small quantities of residual line emission in our quiescent galaxy spectra do not come from ongoing star formation (primarily high [N\,\textsc{ii}]/H$\alpha$ ratios). \textsc{Bagpipes} can only model line emission from ongoing star formation, so we opt to mask the wavelengths of the [O\,\textsc{ii}] 3727\AA\ line and [N\,\textsc{ii}]-H$\alpha$ complex from our fits. We do not observe [O\,\textsc{iii}] 4959,5007\AA\ emission in any of our spectra, and so we do not mask the wavelengths of these lines. We do however also mask the Na\,\textsc{i} 5890\AA,5896\AA\ absorption feature, as this is known to have a strong interstellar medium (ISM) component, which is also not accounted for by \textsc{Bagpipes}.

Whilst we do still include the \textsc{Bagpipes} nebular component in our fits for consistency, the lack of ongoing star formation in our high-redshift quiescent galaxies and the masking of our spectra discussed above mean that the nebular model makes a negligible contribution to the model spectra. We hence do not allow the free parameters of the nebular model to vary in our final run, opting to fix them to representative values for simplicity. The ionization parameter is fixed to log$_{10}(U)=-3$, and the nebular metallicity is constrained to mirror the same value as the stellar metallicity.

We additionally tested fitting these spectra whilst including the AGN component introduced in \cite{Carnall2023a}, which was necessary to obtain a good fit to the spectrum of the $z=4.658$ massive quiescent galaxy, GS-9209, reported in that work. However, none of our new high-redshift quiescent galaxies require this component to obtain a good fit, with the sampler returning the highest probabilities when the normalisation of the AGN component is negligibly small. We therefore remove this component from our final analysis.

To sample the posterior for our joint spectroscopic + photometric fits, we employed the \textsc{Nautilus} code \citep{Lange2023}, now implemented within \textsc{Bagpipes}, which we find to be substantially more efficient than \textsc{MultiNest} for this problem, whilst producing indistinguishable results.

\subsubsection{ALF fitting}\label{subsec:method_alf}

As will be discussed in Section \ref{subsec:results_zf7329}, for one of our galaxies, ZF-UDS-7329, we measure an age of $>1$ Gyr by the process described in Section \ref{subsec:spec_fitting}. This places it in the regime for which individual elemental abundances can be measured with the Absorption Line Fitter (\textsc{Alf}; \citealt{Conroy2012, Conroy2018}) code. We hence run \textsc{Alf} for this object only, in order to supplement the stellar metallicities returned by \textsc{Bagpipes} assuming a fixed, scaled-Solar abundance pattern (see Section \ref{subsec:results_z4_zmet}). The key aim of this analysis is to constrain the level of $\alpha-$enhancement in this galaxy. 

The \textsc{Alf} code makes use of the MILES \citep{Sanchez-Blazquez2006} and extended IRTF \citep{Villaume2017} stellar spectral libraries, as well as the MIST isochrones \citep{Choi2016}. We run \textsc{Alf} in ``simple'' mode, which includes 13 free parameters. These are redshift, stellar age (a single-burst SFH model is assumed), total stellar metallicity $Z_*$, stellar velocity dispersion, and separate abundances for 9 elements including Mg and Fe.

We fit the same rest-frame wavelength intervals described in \cite{Conroy2014}, from $0.40-0.64\,\mu$m and from $0.80-0.88\,\mu$m, additionally masking the potentially ISM contaminated Na\,\textsc{i} feature, as discussed in Section \ref{subsec:spec_fitting}. We place an upper limit on the age of the galaxy at 2 Gyr, which is approximately the age of the Universe at the spectroscopic redshift of $z=3.2$ we measure for this galaxy.

\begin{figure*}
	\includegraphics[width=0.98\textwidth]{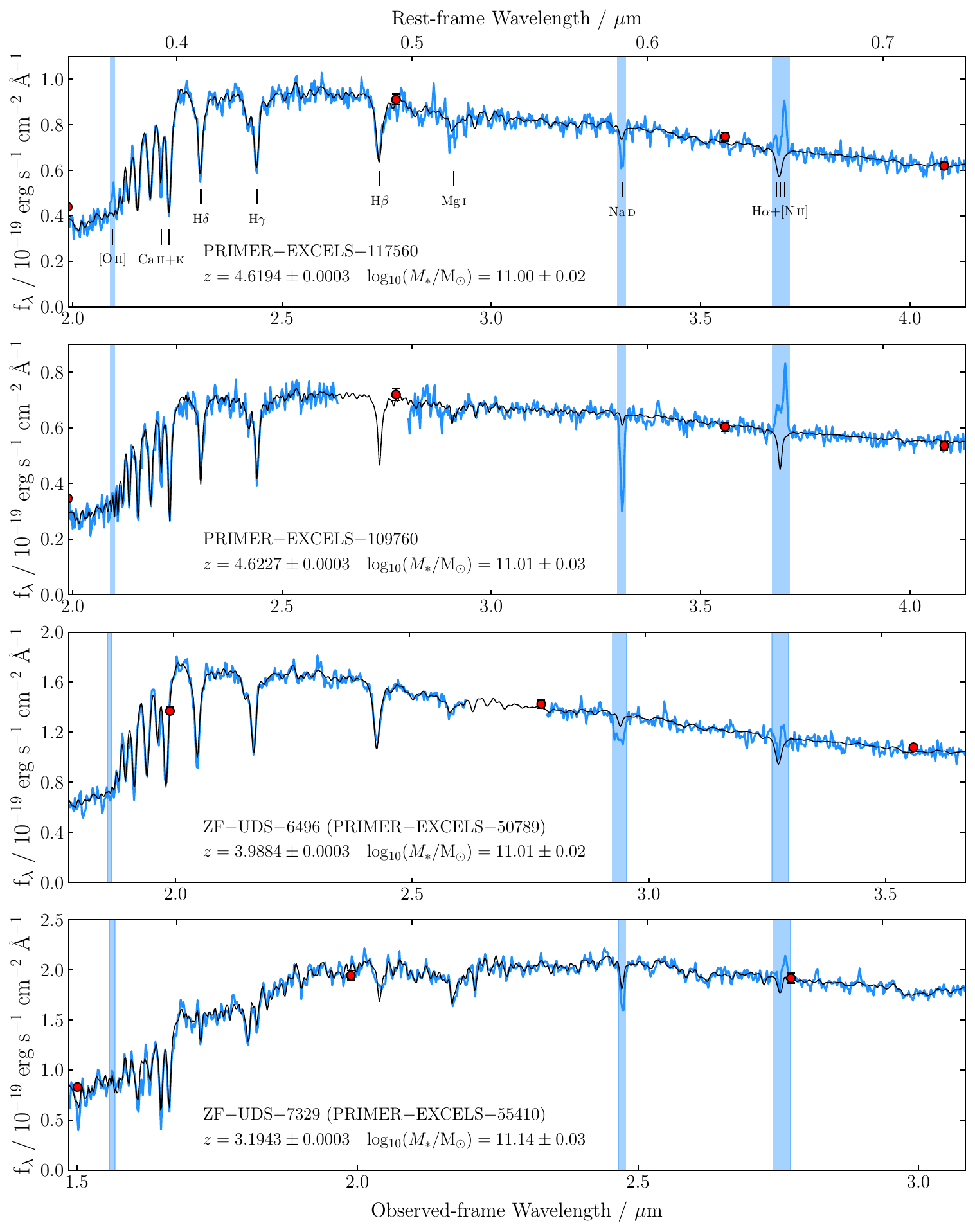}
    \caption{JWST EXCELS NIRSpec observations of our 4 ultra-massive quiescent galaxies at $3 < z < 5$: zoom in on the rest-frame $3540-7350$\AA\ region included in our \textsc{Bagpipes} full-spectral-fitting analysis (see Section \ref{subsec:spec_fitting}). The spectroscopic data are shown in blue, with PRIMER NIRCam photometry shown as red points. The posterior-median fitted \textsc{Bagpipes} models are shown with black lines. The vertical blue shaded regions were masked from the fits. The spectra and our full-spectral-fitting results are described in Section \ref{sec:results}.}
    \label{fig:spec2}
\end{figure*}

\subsection{Size measurement}\label{subsec:size_fitting}

To measure the physical sizes of our ultra-massive quiescent galaxies, we make use of the \textsc{PetroFit} code \citep{Geda2022}, which, in combination with \textsc{Astropy} \citep{Astropy2022}, can be used to model to the observed light distributions of galaxies using S\'ersic profiles. We fit these models to the data using the \textsc{MultiNest} algorithm. 

We fit the light distributions of these galaxies in the PRIMER NIRCam F277W-band data. This is equivalent to rest-frame $\lambda\simeq5000-6000$\AA, just above the Balmer break, and was chosen in order to maximise the SNR with which these objects are detected (they are much fainter blueward of the Balmer break), whilst retaining the highest possible spatial resolution.

We convolve our S\'ersic models with an empirical PSF, determined by stacking bright stars in the field. We fit a $100\times100$ pixel cutout image for each galaxy at the 0.03$^{\prime\prime}$ pixel scale of our mosaic images, whilst manually masking any nearby galaxies. We fit for 8 free parameters: the effective radius $r_e$, S\'ersic index, $n$, normalisation, ellipticity, $e$, position angle, x and y centroids of the S\'ersic profile, as well as the position angle of the PSF. For S\'ersic index we allow values from $0.5-10$ with a uniform prior.

As discussed in Section \ref{subsec:results_sizes}, we have also repeated our analysis on the PRIMER F356W-band imaging, obtaining similar results. We have also repeated this analysis using the \textsc{GalFit} code \citep{Peng2010}, obtaining results consistent to within 10 per cent.

\begin{table*}

\caption{Derived parameters for our 4 EXCELS ultra-massive quiescent galaxies from the \textsc{Bagpipes} full-spectral-fitting analysis described in Section \ref{subsec:spec_fitting}, as well as the morphological analysis described in Section \ref{subsec:size_fitting}. The definitions of the parameters in our full-spectral-fitting analysis are given in Table \ref{table:params}. We also include the results we derived for GS-9209 in \protect\cite{Carnall2023c}. Parameters derived from our \textsc{Bagpipes} full spectral fitting analysis are defined in Table \ref{table:params}. Morphological parameters were measured from F277W-band imaging.}
\begingroup
\setlength{\tabcolsep}{4pt}
\renewcommand{\arraystretch}{1.3}
\begin{tabular}{cccccc}
\hline
Object ID & PRIMER-EXCELS-117560 & PRIMER-EXCELS-109760 & ZF-UDS-6496 & ZF-UDS-7329 & GS-9209 \\

\hline

Redshift & $4.6194\pm0.0003$ & $4.6227\pm0.0003$ & $3.9884\pm0.0003$ & $3.1943\pm0.0003$ & $4.6582\pm0.0002$\\

log$_{10}(M_*/\mathrm{M_\odot})$ & $11.00\pm0.02$ & $11.01\pm0.03$ & $11.01\pm0.02$ & $11.14\pm0.03$ & $10.58\pm0.02$ \\

SFR / M$_\mathrm{\odot}$ yr$^{-1}$ & $0^{+0.0001}_{-0}$ & $0^{+0.004}_{-0}$ & $0^{+0.000001}_{-0}$ & $0.6\pm0.3$ & $0^{+0.000003}_{-0}$ \\ 

log$_{10}(Z_*/\mathrm{Z_\odot})$ & $0.35^{+0.08}_{-0.06}$ & $-0.41^{+0.06}_{-0.09}$ & $0.32^{+0.04}_{-0.05}$ & $0.35^{+0.07}_{-0.08}$ & 
$-0.96^{+0.04}_{-0.09}$\\

$t_\mathrm{form}$ / Gyr & $0.65\pm0.05$ & $0.51\pm0.05$ & $1.01\pm0.03$ & $0.41\pm0.13$ & $0.76\pm0.03$ \\

$z_\mathrm{form}$ & $7.8\pm0.5$ & $9.4\pm0.7$ & $5.6\pm0.1$ & $11.2^{+3.1}_{-2.1}$ & $6.9\pm0.2$ \\

$z_\mathrm{quench}$ & $7.1\pm0.8$ & $6.7\pm0.9$ & $5.4\pm0.2$ & $6.3^{+1.2}_{-1.0}$ & $6.5^{+0.2}_{-0.5}$ \\

$A_V$ & $0.38\pm0.06$ & $0.84\pm0.09$ & $0.49\pm0.05$ & $0.23\pm0.07$ & $0.02\pm0.02$ \\

$a$ & $1.61\pm0.03$ & $1.57\pm0.04$ & $1.57\pm0.04$ & $1.55\pm0.04$ & $1.71\pm0.03$ \\

$\sigma_*$ / km s$^{-1}$ & $360\pm20$ & $140\pm10$ & $370\pm10$ & $250\pm20$ & $250\pm20$ \\

$r_e$ / pc & $610\pm10$ & $310\pm10$ & $730\pm10$ & $910\pm10$ & $220\pm20$ \\

$n$ & $4.7\pm0.1$ & $5.1\pm0.2$ & $3.7\pm0.2$ & $2.5\pm0.1$ & $2.3\pm0.3$ \\

$e$ & $0.63\pm0.01$ & $0.32\pm0.01$ & $0.50\pm0.01$ & $0.77\pm0.01$ & $0.58\pm0.01$ \\

log$_{10}(\Sigma_\mathrm{eff}\ /\ \mathrm{M_\odot}\ \mathrm{kpc}^{-2}$) & $10.63\pm0.03$ & $11.23\pm0.04$ & $10.49\pm0.03$ & $10.42\pm0.04$ & $11.1\pm0.1$ \\

log$_{10}(M_\mathrm{dyn, eff}\ /\ \mathrm{M_\odot}$) & $10.86\pm0.05$ & $9.71\pm0.07$
 & $10.71\pm0.02$ & $10.94\pm0.07$ & $10.3\pm0.1$ \\

\hline
\end{tabular}
\endgroup
\label{table:galaxies}
\end{table*}

\section{Results}\label{sec:results}

The JWST EXCELS dataset provides a wide range of opportunities for studying the formation and evolution of galaxies from cosmic noon back to the first billion years. Having presented both our selection process and the basic demographic properties of the sample in Sections \ref{sec:excels_sample} and \ref{sec:methods} (see Table \ref{table:redshifts} and Fig. \ref{fig:z_vs_mass}), we will move on to exploit this dataset, beginning in this work and continuing in a series of upcoming papers (e.g., Cullen et al. in prep.).

In this section, we present the first set of headline results from EXCELS: the physical properties of the 4 quiescent galaxies we have observed at $3 < z < 5$ for which we derive stellar masses of log$_{10}(M_*/\mathrm{M_\odot}) > 11$ (see Fig. \ref{fig:z_vs_mass}). The SEDs of these galaxies from our PRIMER photometric catalogue (see Section \ref{subsec:primer}) are shown in Fig. \ref{fig:spec1}, and their EXCELS spectra are shown in Fig. \ref{fig:spec2}.

We consider this sub-sample in particular as these galaxies are the most likely to present a problem (by being too early and too massive) for current galaxy formation models and/or $\Lambda$-CDM cosmology. We focus on this issue in this paper, before moving on to a more comprehensive analysis of the whole EXCELS $z>3$ quiescent sample in a follow-up paper.

The EXCELS IDs of these 4 objects in Table \ref{table:redshifts} are 117560, 109760, 50789 and 55410. However, the latter two galaxies have been discussed extensively in several recent papers \citep{Schreiber2018,Nanayakkara2024,Glazebrook2023}, and so, to avoid confusion, for the remainder of this paper we adopt the naming conventions used in these works: ZF-UDS-6496 and ZF-UDS-7329 respectively.

We summarise the key physical parameters we infer for these 4 galaxies in Table \ref{table:galaxies}. We discuss these inferred properties in this section, before moving on in Section \ref{sec:discussion} to discuss whether the SFHs we infer for these most-massive objects place them in tension with the $\Lambda$-CDM halo-mass function.

\begin{figure*}
	\includegraphics[width=\textwidth]{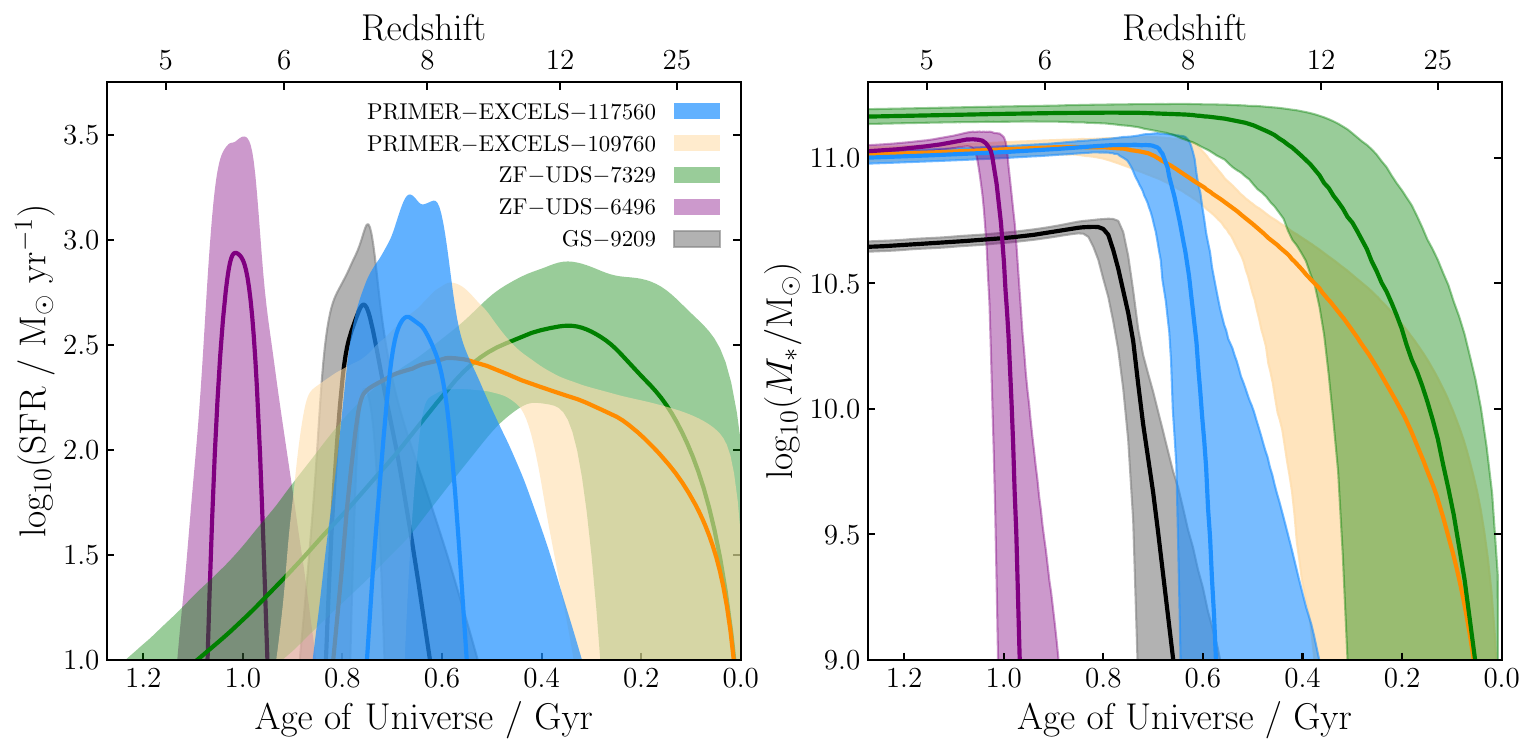}
    \caption{Star-formation histories for our 4 ultra-massive quiescent galaxies at $3 < z < 5$ from full spectral fitting. To the left the SFR as a function of time is shown, whereas to the right the total mass in stars as a function of time is shown. Results for GS-9209 \protect\citep{Carnall2023c} at $z=4.658$, which is $\simeq0.4-0.5$ dex less massive than the other galaxies, are also shown in grey. Three of the new galaxies are older than GS-9209, having formed at $z\gtrsim8$, whereas ZF-UDS-6946 is younger, having formed in a very rapid burst at $z\simeq5.5$. It is instructive to view the shaded areas as confidence intervals on SFR and stellar mass at fixed redshift (i.e., in the vertical direction). Taking PRIMER-EXCELS-109760 as an example, stellar masses of both log$_{10}(M_*/\mathrm{M_\odot})\simeq10.5$ and log$_{10}(M_*/\mathrm{M_\odot})<9$ are within the 1$\sigma$ contour at $z=12$. The right panel therefore indicates we have virtually no constraint on the stellar mass of this galaxy before $z\simeq10$.}
    \label{fig:sfh}
\end{figure*}

\subsection{A pair of massive quiescent galaxies at $\mathbf{z=4.62}$}

The spectra of the two highest-redshift massive quiescent galaxies observed as part of EXCELS are shown in the top two panels of Fig. \ref{fig:spec1}. PRIMER-EXCELS-117560 (top panel) was selected as a robust candidate ($>95$ per cent chance of being at $z>2$ and quiescent) by the process described in Section \ref{subsec:primer_passive}. This galaxy was also selected as a non-robust candidate ($50-95$ per cent chance of being at $z>2$ and quiescent) in our earlier pre-JWST work \citep{Carnall2020} based on the \cite{Galametz2013} catalogue, as well as having been selected independently by \cite{Merlin2019}.

PRIMER-EXCELS-109760 (second panel) was selected as a non-robust candidate (with 86 per cent probability of being at $z>2$  and quiescent) based on the PRIMER NIRCam data, having previously been classified as star-forming ($<50$ per cent chance) in our earlier pre-JWST analysis. 

Both of the spectra for these objects exhibit a forest of extremely deep Balmer absorption features. This is consistent with the spectra of A-type stars, and typical of post-starburst galaxies that have shut down star formation within the past few hundred Myr (e.g., \citealt{Wild2020, DEugenio2020, Werle2022, Wu2023, Leung2024}). As expected based upon this, the SFRs we derive from our full-spectral-fitting analysis (see Table \ref{table:galaxies}) place these galaxies well below the sSFR < 0.2/$t_\mathrm{H}(z)$ threshold we use to define quiescence (see Section \ref{subsec:primer_passive}).

Interestingly, despite a general lack of line emission, both spectra do exhibit weak [N\,\textsc{ii}]$-$H$\alpha$ complexes, with PRIMER-EXCELS-117560 also exhibiting trace amounts of [O\,\textsc{ii}] 3727\AA\ emission (see Fig. \ref{fig:spec2}). For both galaxies, [N\,\textsc{ii}] 6548,6583\AA\ is significantly stronger than H$\alpha$. Line emission excited via irradiation of gas by young massive stars is associated with [N\,\textsc{ii}]/H$\alpha$ ratios significantly less than 1. Higher [N\,\textsc{ii}]/H$\alpha$ ratios are typically associated with alternative excitation mechanisms such as AGN and shocks (e.g., \citealt{Kewley2006}), or post-Asymptotic Giant Branch stars (post-AGB; e.g., \citealt{Binettte1994, Belfiore2016}).

This is similar to the signature observed in GS-9209 \citep{Carnall2023a}, though neither of these new objects exhibits the same broad H$\alpha$ emission component as GS-9209, a clear indication of the presence of an AGN. Similarly elevated [N\,\textsc{ii}]/H$\alpha$ ratios are also common in the spectra of massive quiescent galaxies at the cosmic noon epoch (e.g., \citealt{Belli2017, Newman2018a}).

It is also interesting to note that PRIMER-EXCELS-109760 exhibits extremely deep Na\,\textsc{i} 5890\AA,5896\AA\ (Na D) absorption, evidence for cool gas in the interstellar medium (e.g., \citealt{Belli2023}). Interestingly, this is also the galaxy in our sample with the highest continuum dust attenuation (as measured by $A_V$; see Table \ref{table:galaxies}), which is known to be correlated with the Na D feature (e.g., \citealt{Roberts-Borsani2019}). There is however no clear evidence that this feature is blueshifted, which would indicate the gas is outflowing, as has been observed in young quiescent galaxies at lower redshift (e.g., \citealt{Maltby2019}) and recently at $z=4$ \citep{WuPF2024}. Unfortunately, the Mg\,\textsc{ii} 2800\AA\ ISM absorption feature falls in the NIRSpec chip gap for this object. In future work, we will investigate in detail the line emission and ISM absorption features in the EXCELS quiescent spectra, to assess any potential evidence for ongoing AGN activity. 

It is also interesting to note from Table \ref{table:galaxies} that we obtain consistent values for our spectroscopic errorbar expansion parameter, $a\simeq1.6$, for all 4 of our galaxies. This is in good agreement with the correction factor derived by others (e.g., \citealt{Maseda2023}).

We finally note that PRIMER MIRI photometry is available for both of these galaxies, and we show the F770W fluxes (measured as described in Section \ref{subsec:primer}) in Fig. \ref{fig:spec1}, along with our HST+NIRCam photometry and the posterior median models fitted to these data in Section \ref{subsec:method_specz}. The MIRI photometry (which probes rest-frame $\lambda\simeq1.4\mu$m for these galaxies) can be seen to be in good agreement with the predictions of our fitted models.

\begin{figure*}
	\includegraphics[width=\textwidth]{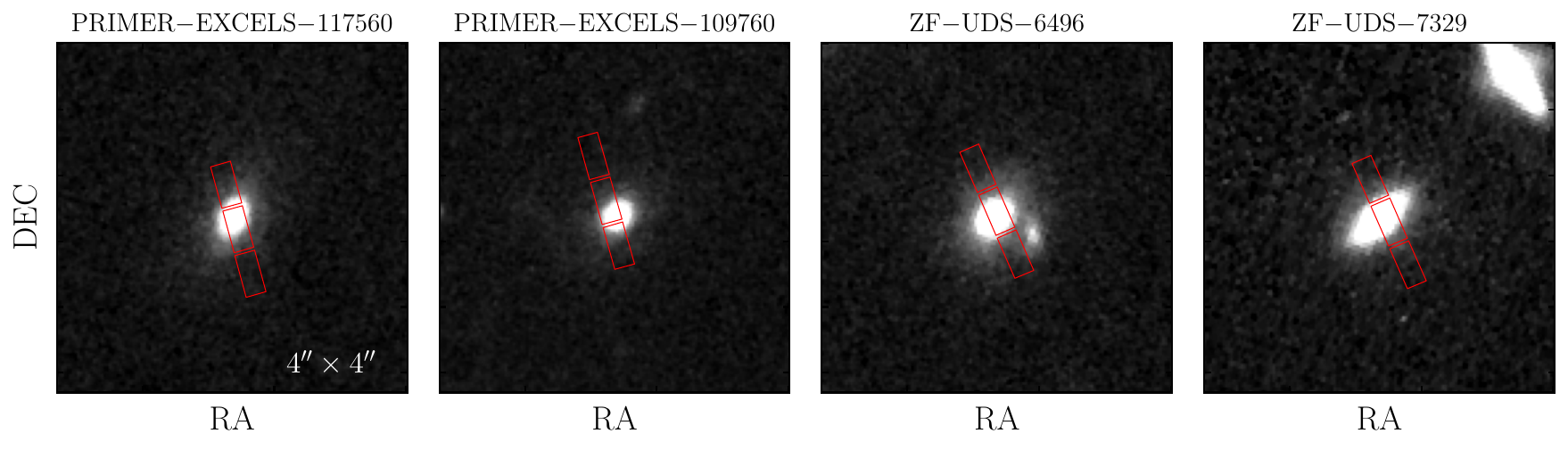}
    \caption{PRIMER F277W cutout images ($4^{\prime\prime}\times4^{\prime\prime}$) for our 4 ultra-massive quiescent galaxies. The positions of the open NIRSpec MSA shutters in the EXCELS G235M observations are shown in red. The positions shown are for the first of three nod positions. Objects were shifted to the top and then bottom shutters for equal thirds of the total exposure time. All four objects are extremely compact ($r_e$ < 1 kpc). PRIMER-EXCELS-117560 and ZF-UDS-7329 appear significantly elongated, whereas the other two objects appear almost round.}
    \label{fig:cutouts}
\end{figure*}

\subsubsection{Star-formation histories}

In addition to our spectroscopic data, in Fig. \ref{fig:spec2} we also show our posterior median \textsc{Bagpipes} models, which were fitted to our data by the process described in Section \ref{subsec:spec_fitting}. We report the precise redshifts and stellar masses that we obtain via our full-spectral-fitting methodology in Table \ref{table:galaxies}. Both $z=4.62$ galaxies have almost identical stellar masses of log$_{10}(M_*/\mathrm{M_\odot})\simeq11$. The SFHs we recover via our full-spectral-fitting methodology are shown in Fig. \ref{fig:sfh}. We also show the SFH derived for GS-9209 in \cite{Carnall2023c}. It can be seen that both of our new $z=4.62$ galaxies are older than GS-9209 (as well as being substantially more massive), with both also exhibiting more-extended SFHs.

We define the time of formation, $t_\mathrm{form}$, as the time after the Big Bang at which the 50th percentile of the stellar mass in each galaxy formed. For these two galaxies at log$_{10}(M_*/\mathrm{M_\odot})\simeq11$, this is equivalent to the time at which they reached log$_{10}(M_*/\mathrm{M_\odot})\simeq10.7$. Using this definition, for PRIMER-EXCELS-117560 we measure $t_\mathrm{form}~=~0.65~\pm~0.05$ Gyr, equivalent to a formation redshift of $z_\mathrm{form}~=~7.8~\pm~0.5$. For PRIMER-EXCELS-109760, we measure an earlier $t_\mathrm{form}~=~0.51~\pm~0.05$ Gyr, equivalent to a formation redshift of $z_\mathrm{form}~=~9.4~\pm~0.7$. 

We also define the time of quenching, $t_\mathrm{quench}$, as the time after the Big Bang at which the galaxy first satisfied the sSFR~<~0.2/$t_
\mathrm{H}(z)$ criterion set out in Section \ref{subsec:primer_passive}. For PRIMER-EXCELS-117560 we measure $t_\mathrm{quench}~=~0.74~\pm~0.10$ Gyr, equivalent to a quenching redshift of $z_\mathrm{quench}~=~7.1~\pm~0.8$. As can be seen from Fig.~\ref{fig:sfh}, for PRIMER-EXCELS-109760 we measure a slightly later time of quenching, $t_\mathrm{quench}~=~0.80~\pm~0.13$ Gyr, equivalent to a quenching redshift of $z_\mathrm{quench}~=~6.7~\pm~0.9$.

It should be noted that the simple parametric double-power-law prior we use to model the SFHs of these galaxies does not contain any physical information about the likelihood of extremely early galaxy formation (see Section \ref{subsec:disc_sfhs}). As can be seen in Fig. \ref{fig:sfh}, the spectrum of PRIMER-EXCELS-109760 is consistent with significant star formation having taken place very early, at $z\gtrsim12$. It is however critical to note that this is not \textit{required} to explain the data. Little to no star formation before $z=11$ can also be seen to be consistent with the data at the $1\sigma$ level. This will be discussed further in Section \ref{sec:discussion}.

\subsubsection{Stellar metallicities}\label{subsec:results_z4_zmet}

Meaningfully measuring the stellar metallicities of high-redshift quiescent galaxies is very challenging. This is because such objects contain relatively young stellar populations (typically $\lesssim1$ Gyr), which exhibit only weak metal absorption features compared with older stellar populations. Another significant issue is that these galaxies are expected to be highly $\alpha-$enhanced (e.g., \citealt{Kriek2016, Carnall2022b, Beverage2023}). However, well established, thoroughly tested, empirical $\alpha-$enhanced stellar population models are not currently available for ages below 1 Gyr.

In the absence of such models, we have used the 2016 updated version of the \cite{Bruzual2003} models, which assume scaled-Solar abundances (i.e., all elemental abundances are multiplied by the same factor with respect to their Solar abundance). In this work, we report the scaled-Solar abundances returned by \textsc{Bagpipes} using the \cite{Bruzual2003} models, whilst cautioning that their interpretation is less than fully clear given the expected $\alpha-$enhancement of our target galaxies. In future work, we will explore extending our analysis to include recently published $\alpha-$enhanced stellar-population models (e.g., \citealt{Knowles2021, Knowles2023, Byrne2022}).

For PRIMER-EXCELS-117560, our full-spectral-fitting analysis returns a stellar metallicity of log$_{10}(Z_*/\mathrm{Z_\odot})=0.35^{+0.08}_{-0.06}$. In marked contrast to this, we recover a significantly lower stellar metallicity for PRIMER-EXCELS-109760 of log$_{10}(Z_*/\mathrm{Z_\odot})=-0.41^{+0.06}_{-0.09}$. This slightly puzzling result appears to be qualitatively well supported by a simple visual inspection of the two spectra in Fig. \ref{fig:spec2}. Both spectra exhibit the extremely deep Balmer absorption lines associated with $\simeq500$ Myr old stellar populations, however PRIMER-EXCELS-117560 exhibits a much stronger 4000\AA\ break (D$_n4000=1.29\pm0.01$ for PRIMER-EXCELS-117569, whereas D$_n4000=1.21\pm0.02$ for PRIMER-EXCELS-109760). Whilst the 4000\AA\ break is often used as an age indicator, it also has a well-known secondary dependence on stellar metallicity (e.g., \citealt{Beverage2021}). The Mg\,\textsc{i} triplet feature at 5170\AA\ is also far more pronounced in the spectrum of PRIMER-EXCELS-117560. Regardless of the precision of these measurements in the absence of $\alpha-$enhanced models, it seems clear that these two galaxies exhibit very different stellar metallicities. 

Our result for PRIMER-EXCELS-117560 is broadly consistent with the stellar metallicities reported by \cite{Beverage2023} for Heavy Metal survey \citep{Kriek2023} massive quiescent galaxies at $z\simeq1.4$ and $z\simeq2.1$, and we also obtain very similar results for ZF-UDS-6496 and ZF-UDS-7329. The metallicity for PRIMER-EXCELS-109760 is much lower, though much more consistent with the low metallicity we measured for GS-9209 in \cite{Carnall2023c}.

Whilst unexpected, this result is not without precedent, given that ultra-massive quiescent galaxies at $z\simeq2$ have also been reported to exhibit a broad spread in stellar metallicity (e.g., \citealt{Kriek2016, Jafariyazani2020}). This result hints at significantly different evolutionary pathways for our two galaxies, despite their sharing several of the same basic properties (e.g., stellar mass) and having formed in relatively close proximity (see Section \ref{subsec:structure}). A potential explanation would be that PRIMER-EXCELS-109760 and GS-9209 are the product of recent mergers, meaning their stars formed in shallower potential wells, hence losing more of their metals to more-efficient outflows of enriched gas.

\subsubsection{Physical sizes}\label{subsec:results_sizes}
We show PRIMER F277W cutout images for our two $z=4.62$ massive quiescent galaxies in Fig. \ref{fig:cutouts}. We measure effective radii, $r_e$, and S\'ersic indices, $n$, following the process described in Section \ref{subsec:size_fitting}. For PRIMER-EXCELS-117560, we obtain $r_e=610\pm10$ pc and $n=4.7\pm0.1$. For PRIMER-EXCELS-109760, we obtain $r_e=330\pm10$ pc and $n=5.2\pm0.2$. These statistical uncertainties are extremely small, as is common for galaxy size measurements (e.g., \citealt{Ji2024}). To gain some estimate of the systematic uncertainty, we apply the same fitting process to the F356W band, obtaining results consistent to within $\simeq10$ per cent.

These measurements place these galaxies significantly below the average size-mass relations for lower-redshift quiescent galaxies of the same stellar mass (e.g., $\simeq2.5$ kpc at $z\simeq1.1$, or $\simeq4$ kpc at $z\simeq0.7$; \citealt{Hamadouche2022}). Our results are however broadly consistent with the $\simeq250-800$ pc sizes measured for other photometrically selected $z>4$ quiescent galaxies by \cite{Ito2023, Wright2023, Ji2024}.

Interestingly, the size of PRIMER-EXCELS-109760 is most similar to GS-9209, and both of these galaxies are smaller by at least a factor of 2 than the other 3 galaxies we analyse in this work. It is possible that this difference could be in some way linked with the apparently substantially lower stellar metallicities of these systems, however our small sample can provide no evidence for this.

We calculate the stellar mass surface density within $r_e$ for these galaxies, obtaining values of log$_{10}(\Sigma_\mathrm{eff}\ /\ \mathrm{M_\odot}\ \mathrm{kpc}^{-2}) = 10.63\pm0.03$ and $11.23\pm0.04$ for PRIMER-EXCELS-117560 and 109760 respectively. This places these objects amongst the most dense stellar systems in the Universe, and consistent with the inner regions of local elliptical galaxies (e.g., \citealt{Hopkins2010}). PRIMER-EXCELS-109760 appears to be particularly extreme, with a stellar mass surface density $\gtrsim0.5$ dex in excess of the other galaxies in our sample, and even exceeding GS-9209, by $\simeq0.1$ dex.

\begin{figure*}
	\includegraphics[width=\textwidth]{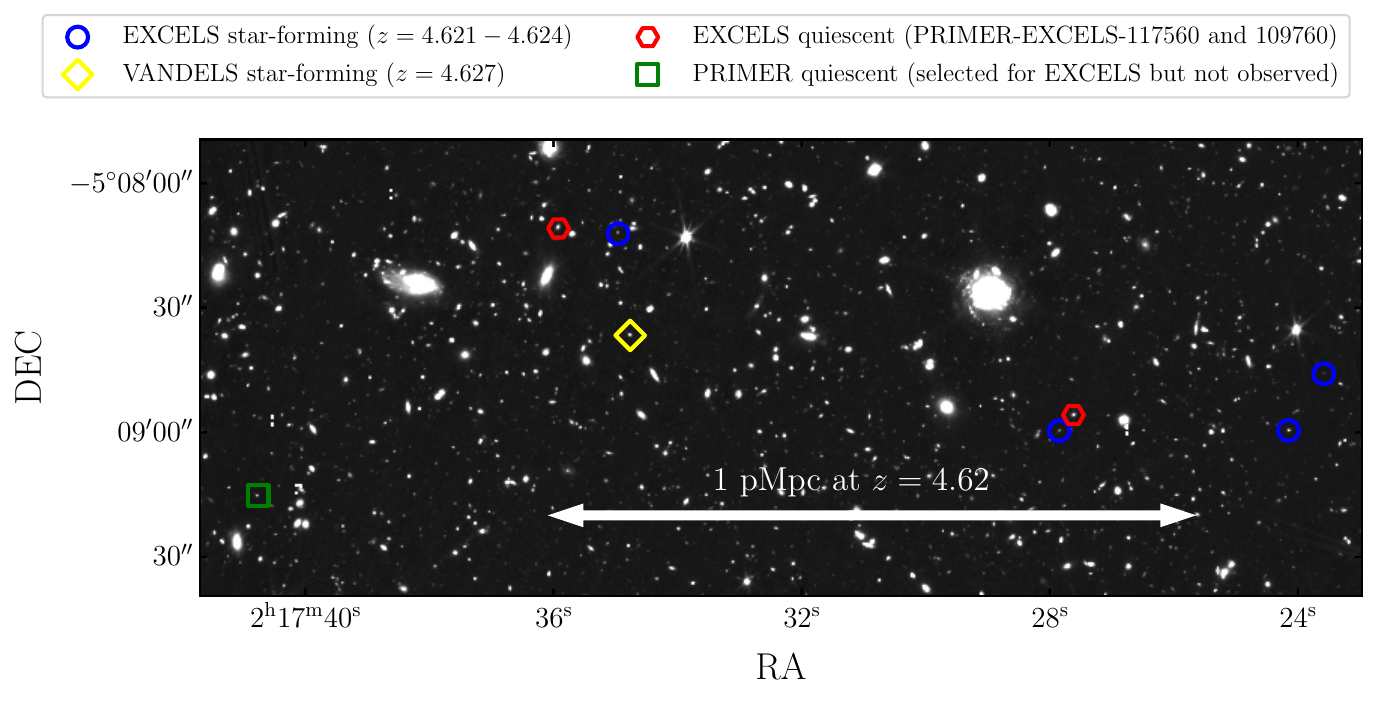}\vspace{-0.2cm}
    \caption{A section of PRIMER F356W imaging, showing spectroscopically confirmed galaxies in the structure we report at $z\simeq4.62$. Red hexagons show our massive quiescent galaxies. Four further EXCELS star-forming galaxies are shown with blue circles, including close ($<100$ kpc) companions for each of the quiescent galaxies. A single VANDELS star-forming galaxy at this redshift is also shown with a yellow diamond. The white arrow shows a proper distance of 1 Mpc at $z=4.62$. We finally show, with a green square, a $z=4.74\pm0.15$ quiescent galaxy candidate, photometrically selected from PRIMER by the same process used to select our two spectroscopically confirmed massive quiescent galaxies (see Section \ref{subsec:primer_passive}). As there are only six such $z>4$ quiescent candidates across the whole of PRIMER UDS, including the three shown here, we speculate that this galaxy is also likely to be associated with the $z=4.62$ structure.}
    \label{fig:field}
\end{figure*}

\subsubsection{Velocity dispersions}\label{subsec:vd_z4}

As can be seen from our observed spectra shown in Fig. \ref{fig:spec2}, PRIMER-EXCELS-117560 has a significantly higher velocity dispersion than PRIMER-EXCELS-109760. After correction for instrumental dispersion (averaging $\sigma\simeq128$ km s$^{-1}$) and the intrinsic dispersion of our stellar models ($\sigma\simeq70$ km s$^{-1}$), we obtain a stellar velocity dispersion, $\sigma_*=360\pm20$ km s$^{-1}$ for PRIMER-EXCELS-117560. For PRIMER-EXCELS-109760, we measure $\sigma_*=140\pm10$ km s$^{-1}$ after correction for the above effects. It should be noted however that this is approaching the velocity resolution of NIRSpec for our chosen instrument mode. For comparison, in \cite{Carnall2023c} we measured $\sigma_*=250\pm20$ km s$^{-1}$ for GS-9209.

The positions of our NIRSpec MSA slitlets with respect to our target galaxies are shown in Fig. \ref{fig:cutouts}. It can be seen that, whilst PRIMER-EXCELS-117560 is well centred, in the case of PRIMER-EXCELS-109760 the slitlet is offset to the side of the galaxy. This complicates the interpretation of the relatively low $\sigma_*$ value we measure for this galaxy. NIRSpec integral-field unit (IFU) observations, as have recently been approved for both GS-9209 and ZF-UDS-7329, would be of great value in clarifying this situation.

For PRIMER-EXCELS-117560, where the slit is well centred, the result we obtain is in good agreement with previous results for similar objects at $z\simeq3$ \citep{Forrest2022}, as well as consistent with local analogues (e.g., \citealt{DAgo2023,Spiniello2024}). We combine our $\sigma_*$ measurement with our effective radius and S\'ersic index measurements from Section \ref{subsec:results_sizes} to estimate the dynamical mass (e.g., see equation 4 of \citealt{Maltby2019}), obtaining log$_{10}(M_\mathrm{dyn}/\mathrm{M_\odot})=10.86\pm0.05$.

This approach assumes that the velocity dispersions we measure are representative of the central velocity dispersions in these galaxies, and also assumes that light and mass trace each other well. This approach also neglects the possibility of rotation in these objects. The higher-resolution IFU data discussed above should open up the possibility of more-sophisticated dynamical analyses (e.g., \citealt{vanderwel2022}).

The stellar mass we infer within $r_e$ is log$_{10}(M_*/\mathrm{M_\odot})=10.70\pm0.02$, meaning that our dynamical mass is fairly consistent with our measured stellar mass at the $\simeq2\sigma$ level. This suggests both that our stellar-mass measurement is reasonable (i.e., not higher than the dynamical mass), and that the central region of the galaxy is stellar-dominated.

For PRIMER-EXCELS-109760, our size and velocity dispersion measurements suggest a dynamical mass within $r_e$ of log$_{10}(M_\mathrm{dyn}/\mathrm{M_\odot})=9.71\pm0.07$, substantially smaller than our measured stellar mass. One possible explanation for this, aside from the caveats listed above, would be that this is a disk galaxy observed face-on. This is consistent with our low measured ellipticity for this galaxy, $e=0.32\pm0.01$, the smallest value of the galaxies in our sample.

\subsubsection{Physical proximity and companion objects}\label{subsec:structure}

Given the redshifts we measure for these two galaxies from our full-spectral-fitting analysis, we calculate that they have a velocity offset of $170\pm20$ km s$^{-1}$. Given their redshift and separation on the sky, we calculate a physical separation of 860 pkpc perpendicular to our line of sight. We have also serendipitously observed, as part of EXCELS, 4 other nearby star-forming galaxies (within $\simeq1$ pMpc) at almost exactly the same redshift. The positions and redshifts of these objects are given in Table \ref{table:redshifts}, where they have IDs 109269, 109360, 112152 and 117855. In addition, we also identify a further star-forming galaxy in close proximity that was not targeted by EXCELS, but which has a VANDELS spectroscopic redshift of $z=4.627$.  The spatial distribution of spectroscopically confirmed $z=4.62$ objects within $\simeq1$ pMpc of our two massive quiescent galaxies is shown in Fig. \ref{fig:field}. 

Furthermore, the only $z>4$ PRIMER UDS quiescent candidate identified in Section \ref{subsec:primer_passive} that was not observed as part of EXCELS is also $<1$ pMpc from PRIMER-EXCELS-117560. This object has an ID of 106450 in the PRIMER UDS catalogue described in Section \ref{subsec:primer}, with a photometric redshift of $z=4.74\pm0.15$ and a stellar mass of log$_{10}(M_*/\mathrm{M_\odot})=10.12\pm0.04$. We assign it a 68 per cent probability of being quiescent. Given that only 6 such quiescent candidates at $z>4$ were identified across the whole of PRIMER UDS (including the 3 shown in Fig. \ref{fig:field}), we speculate that this quiescent candidate is also likely to be associated with the $z=4.62$ structure.

Whilst it is very challenging to identify structures using only photometric redshifts, due to the large intrinsic uncertainties, we do observe a peak in the photometric redshift histogram for our PRIMER UDS catalogue at $z\simeq4.4-4.8$. This suggests that a significant number of further galaxies are likely to also be associated with this structure, in addition to the 7 spectroscopically confirmed members.

The 5 star-forming galaxies in the structure have stellar masses in the range log$_{10}(M_*/\mathrm{M_\odot})\simeq9.0-9.8$, approximately $1-5$ per cent of the masses of the quiescent galaxies. Two of these low-mass star-forming galaxies are each close companions of the massive quiescent galaxies (117855 is 94 pkpc from 117560 and 109360 is 33 pkpc from 109760). This suggests that the growth of massive quiescent galaxies via minor mergers, well known at lower redshift (e.g., \citealt{McLure2013, Hamadouche2022, Suess2023}), is already beginning to act on these galaxies by $z=4.62$.

Assuming that our $z=4.62$ quiescent galaxies are moving towards each other at a conservative fiducial relative velocity of $\simeq100$ km s$^{-1}$ perpendicular to our line of sight, the timescale for crossing this distance is $\lesssim10$ Gyr. It therefore seems plausible that these objects will interact by $z=0$, potentially undergoing a major-merger event to become one of the most massive galaxies in the present-day Universe.

In the local Universe, a large majority of the most massive quiescent galaxies (log$_{10}(M_*/\mathrm{M_\odot})\gtrsim11.5$) are slow rotators (e.g., \citealt{Emsellem2007, Emsellem2011}), whereas lower-mass quiescent galaxies are predominantly fast rotating. This has been explained as a consequence of the most massive galaxies forming via major mergers between predominantly gas-poor galaxies \citep{Khochfar2003, Khochfar2011}, whereas fast rotators experienced fewer major mergers and accreted a lower fraction of their mass. More-recent work supports this picture, with ultra-massive quiescent galaxies at $z\simeq2-3$ having been found to be fast rotating (e.g., \citealt{Toft2017,Newman2018b,dEugenio2023}).

Assuming our two $z=4.62$ quiescent galaxies have not yet been involved in major mergers, and are therefore fast rotators, a future major merger between them at some point between cosmic noon and the present day, resulting in a slow-rotating galaxy, would be fully consistent with this picture.

\subsection{ZF-UDS-6496: an extreme PSB at $\mathbf{z=3.99}$}

The third panel in Fig. \ref{fig:spec2} shows ZF-UDS-6496 (PRIMER-EXCELS-50789), which was selected from our PRIMER catalogue in Section \ref{subsec:primer_passive} with a 99.98 per cent chance of being at $z>2$ and quiescent. This galaxy met virtually all of the criteria to be included in our pre-JWST analysis in \cite{Carnall2020}, however it was removed due to our reduced chi-squared cut (this galaxy was also noted to be fitted with a relatively high chi-squared value by \citealt{Schreiber2018}).

This galaxy was first spectroscopically observed with Keck/MOSFIRE by \cite{Schreiber2018}, who reported a non-detection, from which they inferred a lack of strong emission lines suggesting this galaxy was in fact quiescent. Spectroscopic confirmation was eventually provided by \cite{Nanayakkara2024} using the JWST NIRSpec prism.

In our medium-resolution data, this galaxy exhibits a very extreme PSB spectral shape, with deep and broad Balmer absorption lines. A summary of the physical properties we infer for this galaxy is given in Table \ref{table:galaxies}. We again infer a mass of log$_{10}(M_*/\mathrm{M_\odot})\simeq11$ for this galaxy, as well as a very similar, relatively high stellar metallicity to PRIMER-EXCELS-117560 of log$_{10}(Z_*/\mathrm{Z_\odot})=0.32^{+0.04}_{-0.05}$.

The SFH we infer for this galaxy is shown in Fig. \ref{fig:sfh}. It can be seen that ZF-UDS-6496 is by far the youngest galaxy in our sample, having formed in an extremely intense, brief starburst event at $z\simeq5.5$. This is slightly younger than the $z_\mathrm{form}=6.1$ reported for this object by \cite{Nanayakkara2024} based on their prism spectrum. We infer a peak SFR during this starburst event of $\simeq1000$ M$_\odot$ yr$^{-1}$, similar to the result obtained by \cite{Deugenio2021}, and comparable with the most extreme submillimetre galaxies at these redshifts (e.g., \citealt{Michalowski2017}).

In the PRIMER F277W cutout image shown in Fig. \ref{fig:cutouts}, this object appears fairly rounded and is well centred within our NIRSpec MSA slitlet. It also appears to have a very close, faint companion object, though unfortunately this is outside of our MSA slitlet. From this F277W image, we measure an effective radius for ZF-UDS-6496 of $r_e=730\pm10$ pc. The corrected stellar velocity dispersion we infer from our full spectral fitting is $\sigma_*=370\pm10$ km s$^{-1}$, and we combine these measurements to infer a dynamical mass within $r_e$ of log$_{10}(M_\mathrm{dyn}/\mathrm{M_\odot})=11.07\pm0.04$. This can be compared with our inferred stellar mass within $r_e$ of log$_{10}(M_*/\mathrm{M_\odot})=10.71\pm0.02$, suggesting that, for this galaxy $\simeq40-50$ per cent of the mass within $r_e$ is in the form of stars.

\subsection{ZF-UDS-7329: a fossilised galaxy at $\mathbf{z=3.19}$}\label{subsec:results_zf7329}

The final galaxy in our sample, shown in the bottom panel of Fig. \ref{fig:spec2}, is ZF-UDS-7329 (PRIMER-EXCELS-55410). This object was selected from our PRIMER catalogue with a 99.94 per cent chance of being $z>2$ and quiescent, having been selected as a non-robust candidate in our earlier, pre-JWST analysis. We summarise the physical properties we derive for this galaxy from  EXCELS and PRIMER in Table \ref{table:galaxies}.

This galaxy was also observed with Keck/MOSFIRE by \cite{Schreiber2018}, with continuum flux detected but no clear features from which to measure a spectroscopic redshift. Spectroscopic confirmation was reported by \cite{Nanayakkara2024} from NIRSpec prism data, with this galaxy then being analysed in more detail by \cite{Glazebrook2023}, as discussed in Section \ref{sec:intro}.

As can be seen from Figs \ref{fig:spec1} and \ref{fig:spec2}, this galaxy has a significantly redder spectral shape than the other 3 objects in our sample. This is largely due to its significantly older stellar population, as this object is at a much lower redshift whilst still having formed very early on in cosmic history.

We derive a stellar mass of log$_{10}(M_*/\mathrm{M_\odot})=11.14\pm0.03$ for this object, slightly lower than (but consistent with) the value of log$_{10}(M_*/\mathrm{M_\odot})=11.26^{+0.03}_{-0.16}$ reported by \cite{Glazebrook2023}. The SFH we derive for this galaxy from full spectral fitting of our EXCELS data is shown in Fig. \ref{fig:sfh}. We derive a formation redshift of $z_\mathrm{form}=11.2^{+3.1}_{-2.1}$ for this galaxy, which again is consistent with the value of $z_\mathrm{form}=10.4^{+4.0}_{-2.2}$ reported by \cite{Glazebrook2023}. 

We do however find that, whilst this galaxy formed the bulk of its stellar mass very early, it also had a much more extended formation epoch than the other 3 galaxies in our sample, continuing to form stars through most of the first billion years of cosmic history, before quenching at $z_\mathrm{quench}=6.3^{+1.3}_{-1.0}$ (however see Section \ref{subsec:disc_sfhs} for potential caveats to this).

We also derive a stellar metallicity, log$_{10}(Z_*/\mathrm{Z_\odot})=0.35^{+0.07}_{-0.08}$, very similar to the values we derive for PRIMER-EXCELS-117560 and ZF-UDS-6496. Because this galaxy is so old, we are also able to fit its observed spectrum with the \textsc{Alf} code to constrain individual elemental abundances, in particular the $\alpha-$enhancement (see Section \ref{subsec:method_alf}). Using \textsc{Alf}, we measure [Fe/H] = $-0.10^{+0.13}_{-0.17}$ and [Mg/Fe] = $0.42^{+0.19}_{-0.17}$. These abundances are commonly converted to a total metallicity assuming log$_{10}(Z_*/\mathrm{Z_\odot})$ = [Z/H] = [Fe/H] + 0.94$\times$[Mg/Fe] (e.g., \citealt{Thomas2003,Kriek2019}). Following this, we obtain log$_{10}(Z_*/\mathrm{Z_\odot}) = 0.30^{+0.16}_{-0.18}$ with \textsc{Alf}, in good agreement with the result we have obtained with \textsc{Bagpipes}. 

Our measured $\alpha-$enhancement is consistent with the $\alpha-$enhancements inferred for the most massive quiescent galaxies at $z\simeq2$. For example, \cite{Kriek2016} find [Mg/Fe] = $0.31\pm0.12$ and \cite{Jafariyazani2020} find [Mg/Fe] = $0.51\pm0.05$. We further discuss the results of our \textsc{Alf} fitting for this object in Appendix \ref{app2}.

As well as being the most-massive galaxy in our sample, it is also the largest, with $r_e=910\pm10$ pc. It can also be seen from Fig. \ref{fig:cutouts} that this object is significantly elongated, which could suggest a disk-like morphology. We derive a stellar velocity dispersion of $\sigma_*=250\pm20$ km s$^{-1}$ for this galaxy, which we combine with our $r_e$ and $n$ measurements to derive a dynamical mass within $r_e$ of log$_{10}(M_\mathrm{dyn}/\mathrm{M_\odot})=10.94\pm0.07$. The stellar mass we measure within $r_e$ is log$_{10}(M_*/\mathrm{M_\odot})=10.84\pm0.03$, suggesting that, as with PRIMER-EXCELS-117560, the inner region of this galaxy is stellar-dominated.

Overall, our analysis of the higher-resolution EXCELS spectroscopic data broadly supports the physical properties derived for this galaxy by \cite{Nanayakkara2024} and \cite{Glazebrook2023} using lower-resolution NIRSpec prism data. We now move on in Section \ref{sec:discussion} to consider their claim that the SFH of this galaxy is inconsistent with the $\Lambda$-CDM halo-mass function.

\section{Discussion}\label{sec:discussion}

The $\Lambda$-CDM cosmological model places an upper limit on the number density of massive galaxies as a function of cosmic time. This is because the number density of sufficiently massive halos must be high enough to accommodate the massive galaxies that exist at a certain redshift (e.g., \citealt{Behroozi2018, Wechsler2018}). Whilst this basic concept is fairly straightforward, in practice making a robust comparison between galaxy stellar masses and the likely masses of available dark matter halos in a certain volume is very challenging, due to the effects of cosmic variance, stochastic sampling of the halo-mass and galaxy-stellar-mass functions, and the limited volumes of observational surveys and simulations. 

A common approach taken by observers is to consider a galaxy (or a grouping of galaxies) with a (typically large) observed stellar mass, as well as the number density for galaxies of this kind implied by the survey volume from which it was selected. Such comparisons often take this number density and convert it into a value of the halo mass for the galaxy, $M_h$, using single fixed parameterisation of the halo-mass function.

The halo mass can then be multiplied by the cosmic baryon fraction ($f_b=0.16$), and then a further factor, the stellar fraction, $f_*$ (often also denoted as $\epsilon$), describing the fraction of the available baryons that have been converted into stars. This gives an implied stellar mass for the galaxy, or, if assuming $f_*=1$, an upper limit on the stellar mass allowed by $\Lambda$-CDM. This corresponds to the case in which all baryons available to the galaxy have been converted into stars. If the observed galaxy stellar mass is larger than this limit, the galaxy comes under suspicion as being in tension with $\Lambda$-CDM (often referred to as the ``impossibly early galaxy problem'', e.g., \citealt{Steinhardt2016}).

It should be noted that these limits are insensitive to the stellar initial mass function (IMF), as they simply assume a certain fraction of baryons have been converted to stars, whilst remaining agnostic about the mass distribution of the stars that are formed. The observational measurements of galaxy stellar masses against which these limits are compared are however highly sensitive to the assumed IMF, with the \cite{Salpeter1955} IMF returning stellar masses a factor of $\simeq1.7$ higher than the more typically assumed \cite{Kroupa2001} or \cite{Chabrier2003} IMFs. We use the \cite{Kroupa2001} IMF throughout this paper.

The next step in sophistication is to measure the SFH of an observed galaxy to constrain its stellar mass at earlier times, and compare these results against the higher-redshift halo-mass function (e.g., \citealt{Glazebrook2017, Glazebrook2023, deGraaff2024}). This can be more constraining, as the halo-mass function evolves rapidly with redshift at early times. This is particularly interesting for quiescent galaxies, which typically formed their stellar masses a long time before they are observed, and for which SFHs can be much more reliably inferred than for star-forming objects.

\subsection{The extreme value statistics approach}\label{subsec:disc:evs}

These kinds of comparisons however do not account for several of the subtleties listed above. In an effort to construct a more-robust comparison, we here adopt the extreme value statistics (EVS, e.g., \citealt{Harrison2011}) approach introduced by \cite{Lovell2023}. The EVS approach has several advantages over the approach described above: it mitigates the problem of characterising a precise selection function for population studies, provides a full probability distribution with upper and lower limits, and can more easily accommodate uncertainties in the baryon and stellar fractions.

In simple terms, this approach takes an analytical description of the halo-mass function and mathematically transforms this into an \textit{exact} probability density function (PDF) for the mass of the most-massive halo in some given volume. In this paper, we employ the \cite{Behroozi2013} halo-mass function, computed via the \textsc{hmf} package \citep{Murray2013}.

Crucially, at high redshift the halo-mass function rapidly increases with time. This means that over a typical observational survey area, the most-massive halo is almost guaranteed to be at the low-redshift end of any specified light cone (see section 2.1 and appendix A of \citealt{Lovell2023}). For example, if one wishes to find the most-massive halo at $z>4$ in the PRIMER UDS area, it is only necessary to integrate over a small redshift interval (e.g., $\Delta z=0.2$) above $z=4$, and the resulting PDF will be almost indistinguishable from the result that would be obtained by integrating all the way from $z=4$ to infinity. This allows the expected mass of the most-massive halo (as well as e.g., $1\sigma$ lower and upper limits) to be calculated in relatively narrow redshift bins, then plotted as a function of redshift (e.g., see fig. 2 of \citealt{Lovell2023}).

This probability distribution for the mass of the most-massive halo can be converted into a probability distribution for the stellar mass of the most-massive galaxy by first multiplying by the baryon fraction, then by some assumed form for the distribution of galaxy stellar fractions. We consider two models for the stellar fraction distribution. Firstly, the fiducial model adopted in \cite{Lovell2023}. This is a truncated lognormal across the interval $0 \leq f_* \leq 1$, which takes the form
\begin{equation}\label{eqn:fstar}
f_* = ln\big(N(\mu, \sigma^2)\big),
\end{equation}

\noindent where $N$ denotes the normal distribution, for which parameters of $\mu=e^{-2}=0.135$ and $\sigma=1$ are assumed, based upon a variety of theoretical and observational constraints. The second model we consider is the maximal case, in which all of the available baryons are converted straight into stars, such that $f_*=1$ in all cases. Here, we are effectively just multiplying the PDF for the highest-mass halo by the assumed global baryon fraction of 16 per cent.

We perform these calculations using the \textsc{Evstats} package\footnote{https://github.com/christopherlovell/evstats} provided by \cite{Lovell2023}. We assume the survey area of 160 sq. arcmin covered by our PRIMER UDS catalogue (see Section \ref{subsec:primer_passive}). We compute the EVS PDF for both of our two models for $f_*$ in $\Delta z=0.2$ bins from $z=4$ to $12$.

\begin{figure*}
	\includegraphics[width=0.98\textwidth]{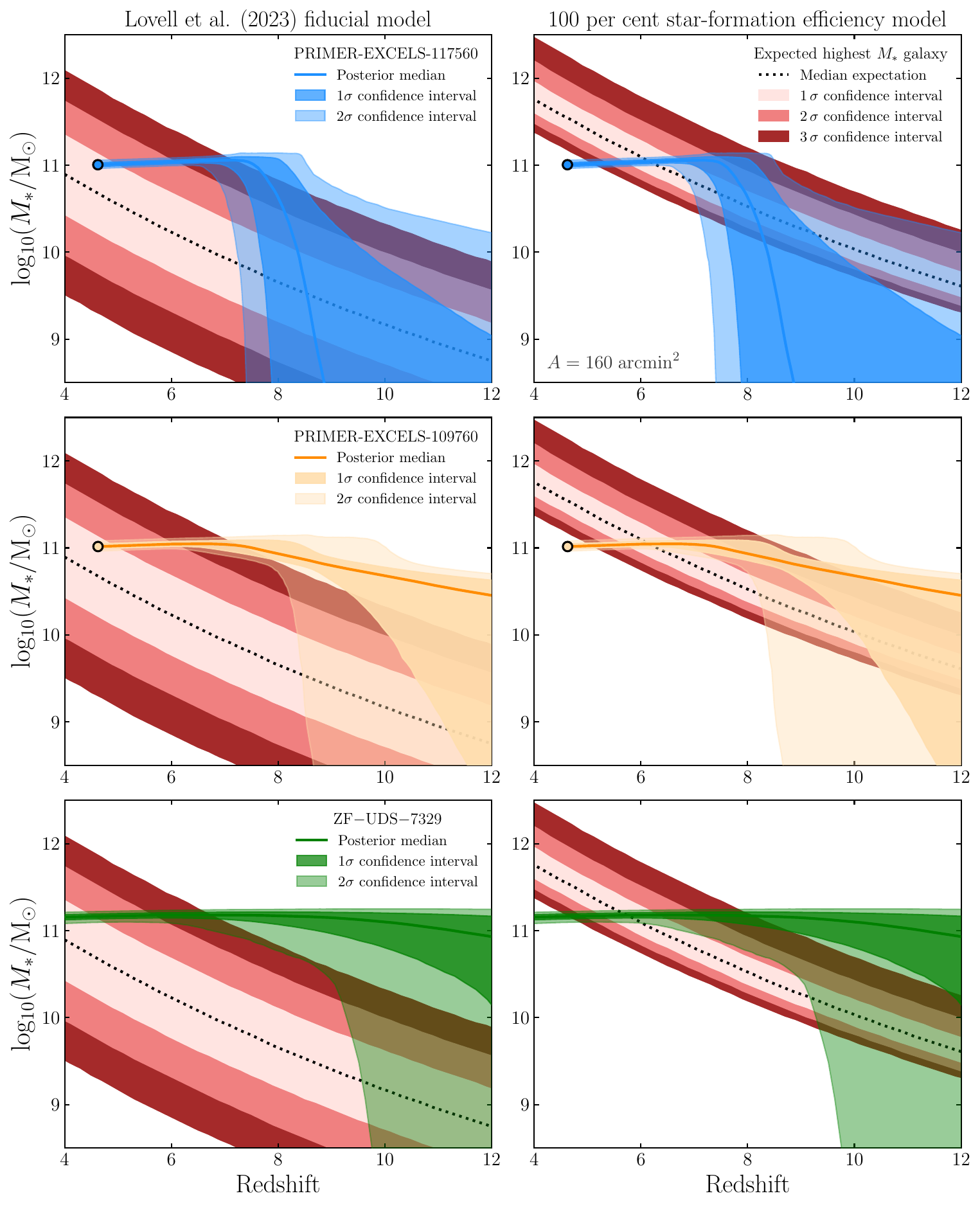}
    \caption{A comparison of the SFHs we derive for our 3 oldest ultra-massive quiescent galaxies with predictions for the most-massive galaxy expected in the PRIMER UDS area as a function of redshift from \protect\cite{Lovell2023}. The extreme value statistics approach used to generate these predictions is discussed in Section \ref{subsec:disc:evs}. To the left, we show the fiducial model presented in \protect\cite{Lovell2023}, which assumes a truncated lognormal distribution of stellar fractions (see Equation \ref{eqn:fstar}). To the right we show our maximum model, which assumes a stellar fraction, $f_*=1$ for all galaxies. The SFHs for all three galaxies are in significant tension with the left-hand model, but can be accommodated by the right-hand model to within a $\lesssim2\sigma$ confidence level. This suggests high stellar fractions for these galaxies.}
    \label{fig:hmf}

\end{figure*}

\subsection{Too much, too young, too fast?}\label{subsec:disc_comparison}

We show our EVS PDFs for the most-massive galaxy expected in our survey volume as a function of redshift in Fig. \ref{fig:hmf}. The left-hand column shows the fiducial model for $f_*$ from \cite{Lovell2023} (see Equation \ref{eqn:fstar}), whereas the right-hand column shows the maximum model, in which $f_*=1$ for all galaxies. We show $1\sigma$, $2\sigma$ and $3\sigma$ contours of the EVS PDF, and denote the median with a dashed black line.

The SFHs we measure for PRIMER-EXCELS-107560, PRIMER-EXCELS-109760 and ZF-UDS-7329 are shown in the top, middle and bottom rows of the figure respectively, with $1\sigma$ and $2\sigma$ confidence intervals shaded. The posterior medians are shown with solid lines. The SFH we measure for ZF-UDS-6496 can be fairly easily accommodated by either of the EVS models shown at all redshifts, and hence we do not show it in this figure.

As with Fig. \ref{fig:sfh}, before a certain point we can place virtually no lower limit on the historical stellar masses of these galaxies (broadly speaking, where the $2\sigma$ lower contours for the SFHs reach the bottom of each panel). For PRIMER-EXCELS-117560 this is around $z\simeq7-8$, for PRIMER-EXCELS-109760 this is around $z\simeq9$, and for ZF-UDS-7329 this is around $z\simeq9-10$. Clearly there can be no conflict with the $\Lambda$-CDM halo-mass function at or before this point, as our observational data for these galaxies are consistent with very low stellar masses.

The period from $z\simeq7-9$, around the time these galaxies quenched their star formation, is however more interesting. It is instructive to focus on the lower contours (i.e., the $2\sigma$ lower limits), which broadly show the latest formation for these galaxies that is consistent with their observed spectra. This is under the assumption of our double-power-law SFH model, which contains no information on the likelihood of extremely early galaxy formation, and simply attempts to bracket the range of SFHs that are consistent with the data (see Section \ref{subsec:disc_sfhs} for further discussion of this point).

It can be seen that, in the left-hand panels, the $2\sigma$ lower limits on our three SFHs cross the $2\sigma$ upper contours for the EVS PDF, and, in the case of PRIMER-EXCELS-109760 and ZF-UDS-7329, even approach the $3\sigma$ upper contour. This suggests that these galaxies are unlikely in our survey volume under the assumption of the \cite{Lovell2023} fiducial model for galaxy stellar fractions (Equation \ref{eqn:fstar}).

In the right-hand panels however, it can be seen that the $2\sigma$ lower contours on our galaxy SFHs are much more consistent with the EVS PDF (at the $\simeq1\sigma$ level). This suggests that, whilst high stellar fractions (approaching $f_*=1$) are required to explain these galaxies around $z\simeq8$, none of them were strongly in tension with the underlying $\Lambda$-CDM halo-mass function at any point in their history.

\subsection{Potential systematic uncertainties}\label{subsec:disc_systematics} 

We here consider several potential sources of systematic uncertainty that could affect the comparison presented in the previous section. We discuss the specific issue of SFH modelling in Section \ref{subsec:disc_sfhs} and Appendix \ref{app1}.

In this analysis, we have considered each of these galaxies separately as the most massive in our survey area at a given redshift. However, it is possible that, for example, ZF-UDS-7329 was the most-massive galaxy at all redshifts, with the other two objects relegated to second and third-most massive. However, given the steepness of the halo-mass function at the high-mass end, and the relatively large volume of our survey, this effect is unlikely to bring any of these galaxies into serious tension with the $\Lambda$-CDM halo-mass function.

Conversely, we have also neglected the potential effects of mergers on these galaxies, which could mean that not all of the stellar mass currently in these galaxies was located within the same halo at higher redshift. It seems unlikely however that all of these galaxies could have undergone major mergers during the relatively short time interval available.

A more top-heavy IMF than the \cite{Kroupa2001} model we have assumed would reduce our implied stellar masses and hence reduce the stellar fractions necessary to accommodate our galaxies. However, the presumed descendants of galaxies such as these in the local Universe actually have more bottom-heavy IMFs than we have assumed in this work (e.g., \citealt{Conroy2012, Maksymowicz-Maciata2024}), which would act to increase the degree of tension. Further discussion of this point can be found in \cite{vanDokkum2024}.

Finally, the impact of cosmic variance on EVS predictions was recently evaluated by  \cite{Jespersen2024}, who demonstrate a wide variation in the predictions of this framework for small volumes. For the relatively large volume probed here however the effect is small, and moves the EVS PDF in the direction of increased tension with measured masses.

\begin{figure*}
	\includegraphics[width=0.8\textwidth]{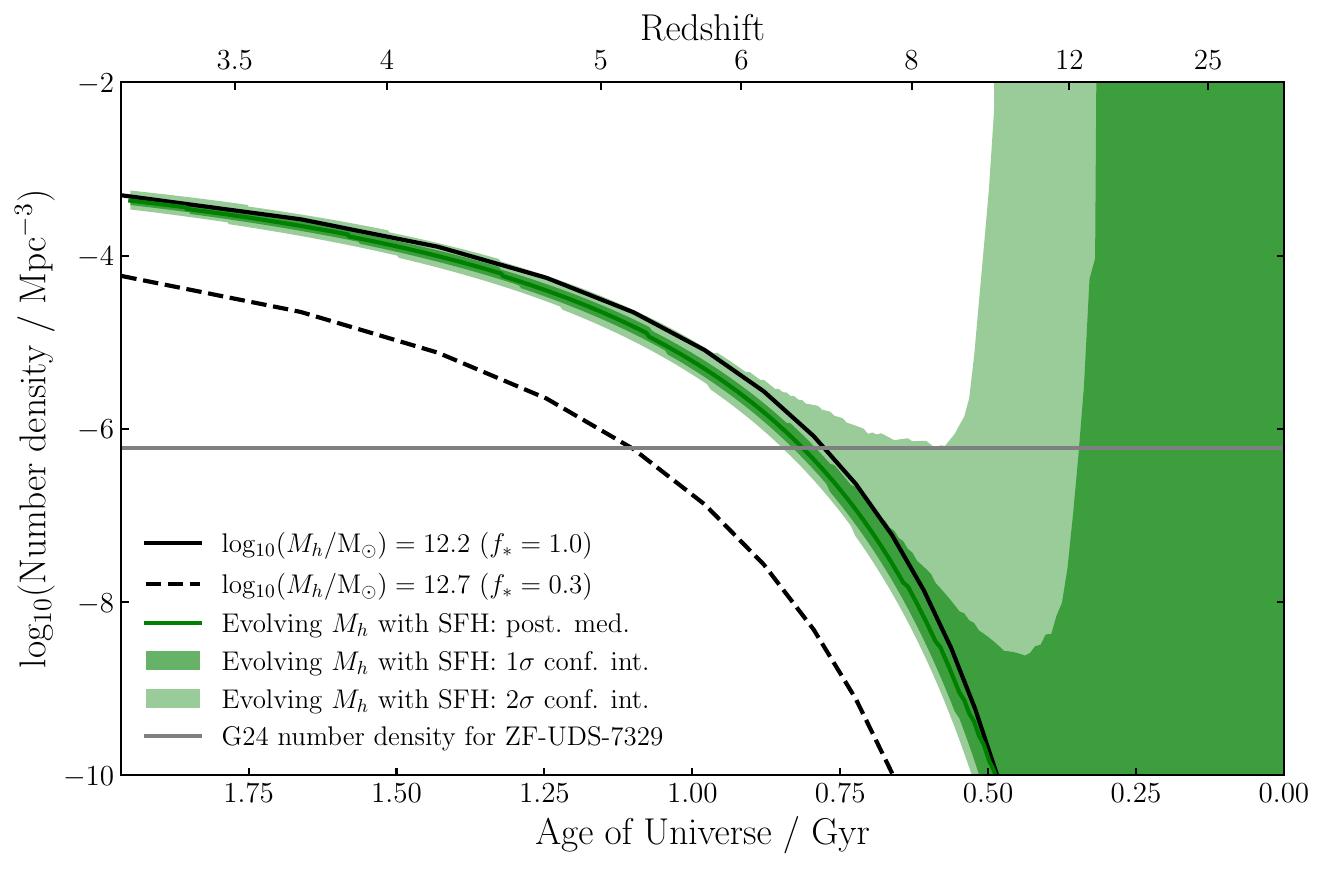}
    \caption{Number densities for the host halo of ZF-UDS-7329 under the assumption of different models for halo mass as a function of redshift (this figure is based on fig. 3 of \protect\citealt{Glazebrook2023}). The black solid and dashed lines are the models presented in \protect\cite{Glazebrook2023}, assuming fixed halo masses of log$_{10}(M_h/\mathrm{M_\odot})=12.2$ and 12.7 respectively at all redshifts. Our model, shown in green, evolves the halo mass down in step with the stellar mass of the galaxy, using the full SFH posterior distributions obtained via full spectral fitting of our EXCELS data (shown in Figs \ref{fig:sfh} and \ref{fig:hmf}). The grey horizontal line is the number density for ZF-UDS-7329 estimated by \protect\cite{Glazebrook2023}. Our model for the halo number density is consistent with the observed number density to within $2\sigma$ at all redshifts.}
    \label{fig:glazebrook_hmf}
\end{figure*}

\subsection{The Glazebrook et al. (2024) halo-mass model}\label{subsec:disc_glazebrook}

The conclusion we present in Section \ref{subsec:disc_comparison}, that the SFH of ZF-UDS-7329 is broadly consistent with the $\Lambda$-CDM halo-mass function (albeit requiring a very high star-formation efficiency), is the opposite of what was concluded by \cite{Glazebrook2023}, despite our finding of a very similar SFH for this object (see Section \ref{subsec:results_zf7329}). In this section we attempt to explain these different conclusions.

In \cite{Glazebrook2023}, the authors take their $z=3.2$ stellar mass for ZF-UDS-7329 and correct for mass loss back into the ISM via supernovae (commonly referred to as the return fraction). They therefore adopt log$_{10}(M_\mathrm{formed}/\mathrm{M_\odot})=11.4$ as the total mass of stars formed by this galaxy by the time we observe it at $z=3.2$. Note that this effect is naturally included in our approach in Section \ref{subsec:disc_comparison}: this is the reason our stellar masses begin to fall at later times in Fig.~\ref{fig:hmf}.

Having computed this total stellar mass formed, they divide through by $f_b=0.16$, and then further divide through by $f_*$, for which they investigate two values: 0.3 and 1. This produces implied halo masses of log$_{10}(M_\mathrm{h}/\mathrm{M_\odot})=12.7$ and 12.2 respectively for this galaxy at $z=3.2$. They then compute the number density of halos with these fixed masses, both at $z=3.2$ and also as a function of redshift back to earlier times, using the \cite{Behroozi2013} halo-mass function. The results of this calculation are shown in their fig. 3.

Crucially, when calculating the expected number density for their ZF-UDS-7329 host halo at higher redshifts, \cite{Glazebrook2023} keep the host halo mass fixed to the values of log$_{10}(M_\mathrm{h}/\mathrm{M_\odot})=12.7$ and 12.2 they calculate at $z=3.2$. It could however be argued that the halo would be expected to grow roughly in step with the galaxy, and would therefore be less massive (and hence more abundant) at higher redshift. For example, by definition, at the formation redshift of the galaxy at $z\simeq11$, the stellar mass is half of its ``final'' value when the galaxy is observed at $z=3.2$. A reasonable assumption therefore might be to also halve the required halo mass when calculating halo number densities at $z\simeq11$, rather than employing the $z=3.2$ value. 

We take a slightly more sophisticated alternative approach by scaling the required halo mass down from the $z=3.2$ value in step with the stellar mass we infer for the galaxy as a function of time, taking into account the full stellar mass posterior distributions shown in Figs \ref{fig:sfh} and \ref{fig:hmf}. We assume a $z=3.2$ halo mass of log$_{10}(M_\mathrm{h}/\mathrm{M_\odot})=12.2$ for consistency with \cite{Glazebrook2023}. 

The results of this calculation are shown in Fig. \ref{fig:glazebrook_hmf} (this figure is based on fig. 3 from \citealt{Glazebrook2023}). We firstly show, with black  dashed and solid lines, our own calculation of the model proposed by \cite{Glazebrook2023}, which assumes fixed halo masses of log$_{10}(M_\mathrm{h}/\mathrm{M_\odot})=12.7$ and 12.2 respectively at all redshifts. In green, we show the posterior median, $1\sigma$ and $2\sigma$ confidence intervals for our model, which begins with log$_{10}(M_\mathrm{h}/\mathrm{M_\odot})=12.2$ at $z=3.2$ and scales this down in line with the SFH we derive for ZF-UDS-7329 from our EXCELS spectroscopic data. It can be seen that the $2\sigma$ upper limit on the number density of the host halo (which is analogous to the $2\sigma$ lower contour on stellar mass shown in Fig. \ref{fig:hmf}) is consistent with the number density inferred for ZF-UDS-7329 by \cite{Glazebrook2023} at all redshifts. As with Fig. \ref{fig:hmf}, there is essentially no constraint before $z\simeq9-10$, because we have virtually no constraint on the stellar mass of the galaxy before this time.

In the end, this comparison is essentially an alternative statement of the one we have developed in Section \ref{subsec:disc_comparison}, and we draw the same conclusion. The $2\sigma$ lower (i.e., later) limit on the SFH we measure for ZF-UDS-7329 is broadly consistent with this galaxy forming within the most-massive halo available in our survey volume, under the assumption of both the $\Lambda$-CDM halo-mass function, and a stellar fraction approaching 100 per cent ($f_*=1$). This conclusion also applies equally to our two new $z=4.62$ quiescent galaxies.

\subsection{The effects of different SFH models}\label{subsec:disc_sfhs}

In our analysis so far, we have assumed a double-power-law SFH model throughout. It is however well known that the SFHs measured for galaxies depend fairly strongly on the assumed form of the SFH model (e.g., \citealt{Carnall2019a,Leja2019a}), though this effect is less pronounced when fitting high-SNR continuum spectroscopy rather than just photometric data (e.g., \citealt{Wild2020}).

We have also focused almost exclusively on the lowest masses allowed by our fitted models as a function of time (e.g., the $2\sigma$ lower limits discussed extensively in Section \ref{subsec:disc_comparison}). There are two key reasons for this. Firstly, it seems clear that it would only be prudent to reject the null hypothesis of the $\Lambda$-CDM halo-mass function on the basis of evidence at much more than a $\lesssim2\sigma$ confidence level. Secondly, almost all SFH models currently in wide usage in the literature were designed with a deliberate bias towards obtaining the \textit{oldest} possible ages for galaxies, typically with a strong focus on the $z<2.5$ Universe.

This choice was made to combat a bias towards very young stellar ages that was discovered to  arise when using the earliest generations of simple SFH models, such as instantaneous burst or exponentially declining SFHs (e.g., \citealt{Wuyts2011}). Such models typically return broadly the youngest ages that are consistent with observed data.

The deliberate bias towards older ages in newer models is particularly true of the current generation of ``non-parametric'' SFHs \citep{Leja2019a,Leja2019c}. Another example is the double-power-law model used in this work, which was designed in \cite{Carnall2018} to accurately reproduce the SFHs of quiescent galaxies at $0.5 < z < 2.5$ in the \textsc{Mufasa} simulation \citep{Dave2016}. Whilst these choices are well motivated within the $z<2.5$ domain for which they were designed, these models have not been extensively tested at, or recalibrated for, the high redshifts at which they are now commonly applied when dealing with JWST data.

In \cite{Glazebrook2023}, the authors also fit their prism spectrum for ZF-UDS-7329 using the continuity non-parametric SFH model of \cite{Leja2019a}, obtaining, as would be expected, an even older age than they obtain with their fiducial dual exponential SFH model, or we obtain with our double-power-law model.

This however is a rare case in which we are actually most interested in the \textit{youngest} age for a galaxy that is consistent with the observed data, rather than the oldest. For this reason, in Appendix \ref{app1}, we repeat our analysis in Section \ref{subsec:disc_comparison} with instantaneous burst SFH models, which are well known to produce something broadly equivalent to a lower limit on the age (as well as the stellar mass) of an observed galaxy, sometimes called the simple stellar population (SSP) equivalent age (e.g., see section 4.2 of \citealt{Conroy2013}). We demonstrate that this change has no effect on the conclusions we present.

More generally, it is concerning that the posterior median SFHs derived for 2 of our ultra-massive quiescent galaxies using our double-power-law model are strongly inconsistent with the $\Lambda$-CDM halo-mass function at $z\gtrsim8$ (PRIMER-EXCELS-109760 and ZF-UDS-7329; see the two lower-right panels of Fig. \ref{fig:hmf}), even if the posterior distributions are consistent to within $\lesssim2\sigma$. This result motivates further thought about whether SFH models that have been developed for the lower-redshift Universe (such as the parametric double-power-law and non-parametric continuity models) can be safely applied to massive galaxies at $z>3$.

\section{Conclusion}\label{sec:conclusion}

In this paper we present the JWST EXCELS survey, a 72-hour Cycle 2 programme targeting quiescent and star-forming galaxies from cosmic noon back to the first billion years for $\lambda=1-5\mu$m medium-resolution NIRSpec spectroscopy. In Section \ref{sec:excels_sample} we present the process by which the 401 EXCELS targets were selected (largely using photometry from the PRIMER Cycle 1 NIRCam programme) and prioritised for space on the NIRSpec MSA. We present a spectroscopic redshift catalogue for EXCELS galaxies in Table \ref{table:redshifts}, and show their stellar mass vs redshift distribution in Fig. \ref{fig:z_vs_mass}.

Headline science results from EXCELS will be published in a series of forthcoming papers (e.g., Cullen et al. in prep.). We begin in this work by analysing the spectra of the 4 ultra-massive (log$_{10}(M_*/\mathrm{M_\odot})>11$) quiescent galaxies at $3 < z < 5$ observed as part of EXCELS. We focus on these objects in particular to address recent debate in the literature about whether the oldest and most massive galaxies at these redshifts are incompatible with current galaxy formation models, and/or the $\Lambda$-CDM halo-mass function.

This sample of 4 objects, shown in Fig. \ref{fig:spec2}, includes: a pair of galaxies at $z=4.62$ (PRIMER-EXCELS-117560 and 109760), physically separated by 860 pkpc within a larger structure for which we spectroscopically confirm an additional 4 members; an extreme PSB galaxy (ZF-UDS-6496) at $z=3.99$ that formed in an intense burst at $z\simeq5.5$; and a relic galaxy at $z=3.19$ (ZF-UDS-7329) that formed the bulk of its stellar mass at $z\simeq11$. We summarise the physical properties we infer for these galaxies in Table \ref{table:galaxies}, and show their SFHs in Fig. \ref{fig:sfh}.

We find a broad range of formation redshifts for these galaxies, from $z\simeq5.5-11$, as well as extremely compact sizes from $r_e\simeq300-900$ pc. We measure typically high (roughly double Solar) stellar metallicities, though PRIMER-EXCELS-109760 appears to exhibit a much lower metallicity, in common with the object GS-9209 we reported in \cite{Carnall2023c}. These two galaxies are also by far the most compact. We are also able to measure the $\alpha-$enhancement for ZF-UDS-7329, obtaining [Mg/Fe] = $0.42^{+0.19}_{-0.17}$. For 3 objects we are able to measure reliable dynamical masses, which suggest high stellar fractions of $\simeq40-90$ per cent in the central regions of these galaxies.

We then consider these galaxies in the context of the $\Lambda$-CDM halo-mass function, using the extreme value statistics approach of \cite{Lovell2023} to calculate expected stellar masses for the most-massive galaxy in the PRIMER UDS field as a function of redshift. We assume two different models for the fraction of the available baryons converted into stars (the stellar fraction, $f_*$): the fiducial model of \cite{Lovell2023} (Equation \ref{eqn:fstar}), which is based on lower-redshift constraints, and a maximum model, which assumes $f_*=1$ for all galaxies.

We present the results of this analysis in Section \ref{subsec:disc_comparison} and Fig. \ref{fig:hmf}. We conclude that 3 of our galaxies are unlikely within the area they were selected under the assumption of standard lower-redshift stellar fractions, but that they can be accommodated by the $\Lambda$-CDM halo-mass function under the assumption of high stellar fractions, approaching $f_*=1$.

This is in contrast to the conclusion recently presented for ZF-UDS-7329 (one of our sample) by \cite{Glazebrook2023}. The SFH we derive for this galaxy is in good agreement with their result, however they conclude that this is in tension with the $\Lambda$-CDM halo-mass function.

In Section \ref{subsec:disc_glazebrook} we compare our model with theirs, which assumes that the halo mass is fixed to its $z=3.2$ value at all earlier times. In Fig. \ref{fig:glazebrook_hmf} we show that, by reducing the required halo mass at earlier times in step with the SFH we derive for this galaxy (e.g., when the stellar mass was half its $z=3.2$ value the halo mass was also halved), the number density of the required halo is consistent with the observed number density for ZF-UDS-7329 to within $2\sigma$ at all redshifts. 

We therefore find no evidence for an ``impossibly early galaxy problem''. However, 3 of our galaxies are at the very edge of what can be accommodated (though, as discussed in Section \ref{subsec:disc_systematics}, several key systematics have the potential to increase or reduce this tension). In this context, our results imply very extreme baryonic physics within the first billion years of cosmic history, unlike anything seen at lower redshift. Trying to understand these objects in more detail must now be a high priority for developing our understanding of early galaxy formation. A key component of this process will be the search for plausible star-forming progenitor objects around the epoch of formation and quenching for our oldest galaxies at $z\simeq8-12$ (e.g., \citealt{Wang2024}).

\section*{Acknowledgements}

A. C. Carnall thanks the Leverhulme Trust for their support via a Leverhulme Early Career Fellowship. A. C. Carnall acknowledges support from a UKRI Frontier Research Guarantee Grant [grant reference EP/Y037065/1]. R. Begley, C. Donnan, D. J. McLeod, R. J. McLure and J. S. Dunlop acknowledge the support of the Science and Technology Facilities Council. F. Cullen, K. Z. Arellano-C\'{o}rdova and T. Stanton acknowledge support from a UKRI Frontier Research Guarantee Grant [grant reference EP/X021025/1]. J. S. Dunlop also acknowledges the support of the Royal Society through a Royal Society Research Professorship. P. Santini acknowledges INAF Mini Grant 2022 ``The evolution of passive galaxies through cosmic time''. Based on observations with the NASA/ESA/CSA James Webb Space Telescope, obtained via the Mikulski Archive for Space Telescopes at the Space Telescope Science Institute, which is operated by the Association of Universities for Research in Astronomy, Incorporated. Support for Program number JWST-GO-03543.014 was provided through a grant from the STScI under NASA contract NAS5-03127.

\section*{Data Availability}

All raw JWST and HST data products are available via the Mikulski Archive for Space Telescopes (\url{https://mast.stsci.edu}). Reduced photometric data and fitted model posteriors are available upon request.

\bsp

\bibliographystyle{mnras}
\bibliography{bibliography} 

\appendix 

\setcounter{figure}{0}
\renewcommand{\thefigure}{A\arabic{figure}}

\section{Further details of \textsc{ALF} spectral fitting of ZF-UDS-7329}\label{app2}

\begin{table}

\caption{Stellar masses, formation times and redshifts for our four ultra-massive quiescent galaxies under the assumption of an instantaneous burst SFH model. This model gives a lower limit on the masses and ages of these objects.}
\begingroup
\setlength{\tabcolsep}{2pt}
\renewcommand{\arraystretch}{1.3}
\begin{center}

\begin{tabular}{lccc}
\hline
ID & log$_{10}(M_*/\mathrm{M_\odot})$ & $t_\mathrm{form}$ / Gyr & $z_\mathrm{form}$ \\
\hline
PRIMER-EXCELS-117560 & $11.00\pm0.02$ & $0.64\pm0.06$ & $7.8\pm0.5$ \\
PRIMER-EXCELS-109760 & $11.01\pm0.03$ & $0.55\pm0.11$ &  $8.8^{+1.6}_{-1.1}$ \\
ZF-UDS-6496 & $11.01\pm0.03$ & $1.02\pm0.04$ & $5.5\pm0.1$ \\
ZF-UDS-7329 & $11.12^{+0.04}_{-0.07}$ & $0.49^{+0.24}_{-0.22}$ & $9.6^{+5.2}_{-2.4}$ \\

\hline
\end{tabular}
\end{center}
\endgroup
\label{table:galaxies_burst}
\end{table}

\begin{figure*}
	\includegraphics[width=\textwidth]{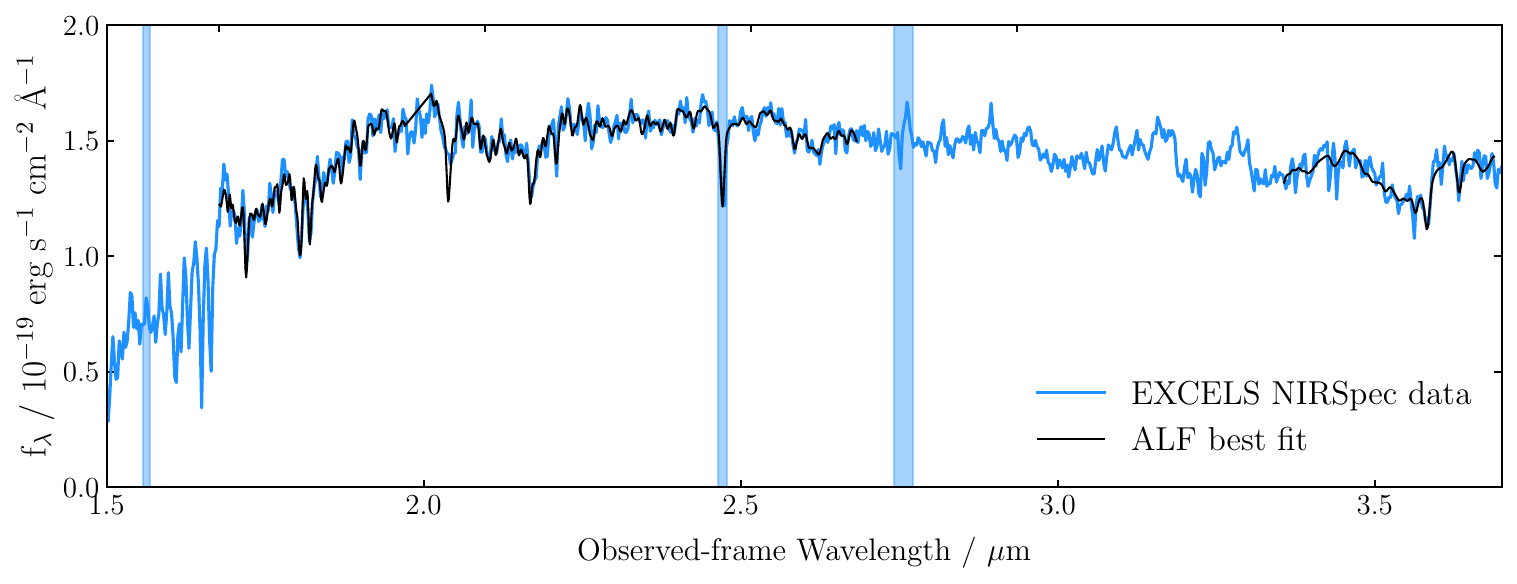}
    \caption{Full spectral fitting of our EXCELS spectroscopic data for ZF-UDS-7329 with the \textsc{Alf} code. Our NIRSpec data are shown in blue (as in Fig. \ref{fig:spec2}), whereas our best-fitting \textsc{Alf} model is shown in black. Fitting was only conducted over the spectral ranges from $0.40-0.64\ \mu$m and $0.80-0.88\ \mu$m following \protect\cite{Conroy2014}. The shaded blue vertical regions are those that were excluded from our \textsc{Bagpipes} fitting, and are also excluded from our \textsc{Alf} fitting where they fall within the above wavelength ranges. Our fitting methodology is described in full in Section \ref{subsec:results_zf7329}. The parameters of our fitted model are reported in Section \ref{subsec:results_zf7329} and Appendix \ref{app2}.}
    \label{fig:alf_spec}

\end{figure*}

We have performed full spectral fitting of our NIRSpec data for ZF-UDS-7329 using the \textsc{Alf} code as described in Section \ref{subsec:method_alf}. The key elemental abundance results derived via this approach are presented in Section \ref{subsec:results_zf7329}; we here present further results from this fitting approach. We show our best-fit \textsc{Alf} model along with our spectroscopic data in Fig. \ref{fig:alf_spec}.

In addition to the iron and magnesium abundances reported in Section \ref{subsec:results_zf7329}, our \textsc{Alf} fitting returns an age of $1.9^{+0.1}_{-0.3}$ Gyr, corresponding to $t_\mathrm{form}=0.1^{+0.3}_{-0.1}$ Gyr. This is broadly consistent with our \textsc{Bagpipes} result of $t_\mathrm{form}=0.41\pm0.13$ to within $1\sigma$.

The $\alpha$ abundance measured from our \textsc{Alf} fitting can be converted into an implied formation timescale using the relationship proposed by \cite{Thomas2005}. We obtain an implied formation timescale of $5^{+496}_{-2}$ Myr. This 1$\sigma$ upper limit is broadly consistent with the formation timescale we derive from our \textsc{Bagpipes} full spectral fitting (e.g., see Fig. \ref{fig:sfh}). The median result is extremely short compared with our \textsc{Bagpipes} fitting, and arguably also with respect to plausible physical timescales for the formation of such a massive galaxy.

A similar result was also recently obtained by \cite{Beverage2024}, who find that the formation timescales implied by their \textsc{Alf}-derived $\alpha$-enhancement values are significantly shorter than the SFHs they derive from \textsc{Prospector} full spectral fitting. This is further explored in \cite{MarcelinaGountanis2024}.

\begin{figure*}
	\includegraphics[width=0.98\textwidth]{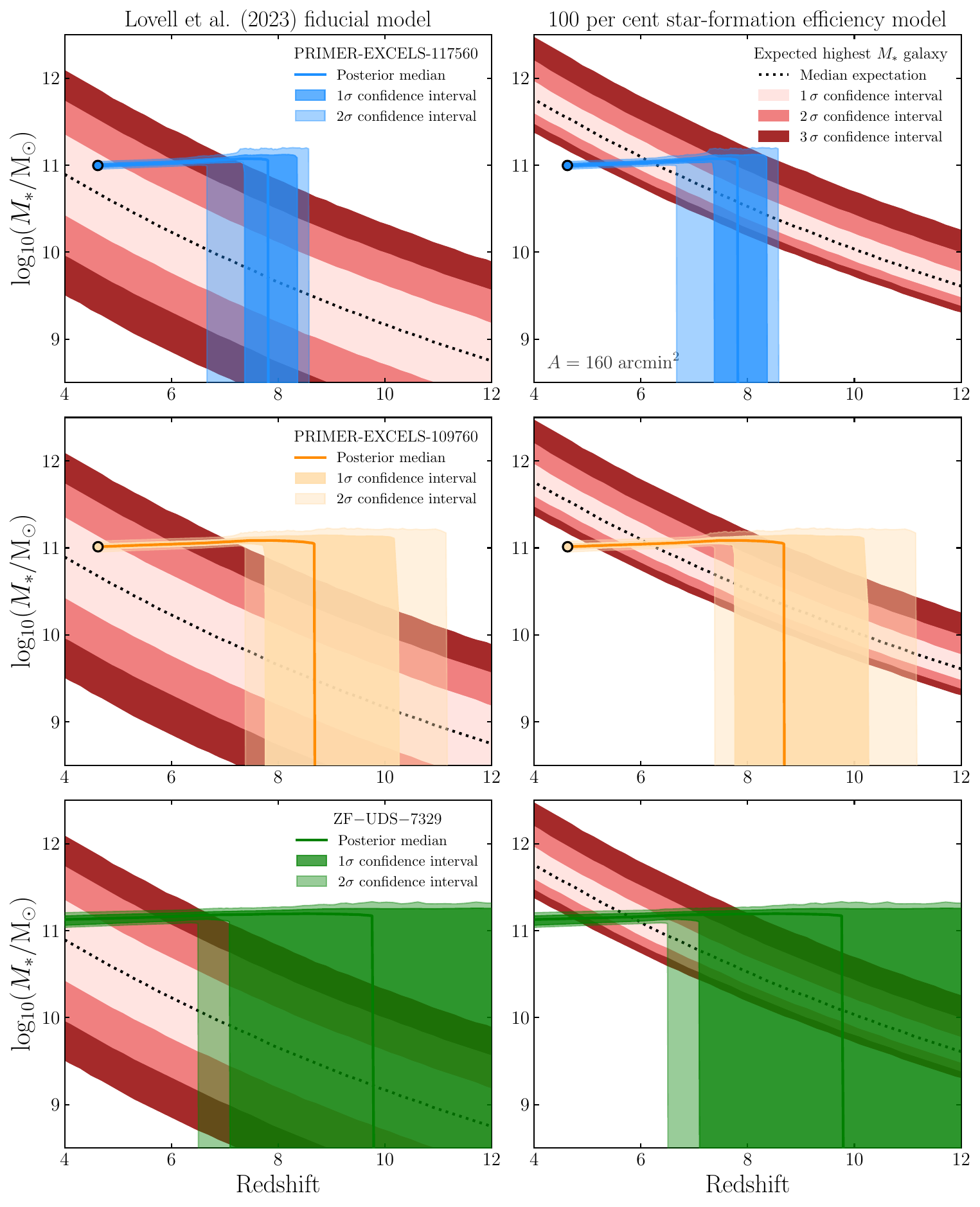}
    \caption{Alternative version of Fig. \ref{fig:hmf}. A comparison of the SFHs we derive for our 3 oldest ultra-massive quiescent galaxies using an instantaneous burst SFH model with predictions for the most-massive galaxy expected in the PRIMER UDS area as a function of redshift from \protect\cite{Lovell2023}. The extreme-value-statistics approach used to generate these predictions is discussed in Section \ref{subsec:disc:evs}. To the left, we show the fiducial model presented in \protect\cite{Lovell2023}, which assumes a truncated lognormal distribution of stellar fractions (see Equation \ref{eqn:fstar}). To the right we show our maximum model, which assumes a stellar fraction, $f_*=1$ for all galaxies. The use of this alternative SFH model to fit our data can be seen to make no changes to the conclusions we draw in Section \ref{sec:discussion} based upon Fig. \ref{fig:hmf}.}
    \label{fig:hmf_burst}

\end{figure*}

\section{Instantaneous burst SFHs}\label{app1}

As discussed in Section \ref{subsec:disc_sfhs}, the results we present are potentially sensitive to the SFH model assumed. Testing alternative SFH models that are designed to give older ages (such as the continuity non-parametric model of \citealt{Leja2019a}) would be of no consequence here, as we have already demonstrated that the younger ages produced by our double-power-law SFH model are consistent with our data. Our conclusion on the need for high stellar fractions could however be overturned if another suitable SFH model could be found that returned significantly younger ages for these galaxies.

To test this, we re-run our full-spectral-fitting analysis, described in Section \ref{subsec:spec_fitting}, using an instantaneous burst SFH model in place of our double-power-law model. As discussed in Section \ref{subsec:disc_sfhs}, this burst model is known to produce lower limits on the ages of galaxies (SSP-equivalent ages). 

The SFHs we obtain from this alternative round of fitting are shown in Fig. \ref{fig:hmf_burst}, which is an alternative version of Fig. \ref{fig:hmf}. It can be seen that the assumption of this alternative SFH model does not substantially change the results reported in Section \ref{sec:discussion}. The instantaneous burst SFHs are still inconsistent with the fiducial model of \cite{Lovell2023} shown in the left-hand panels, whilst the $2\sigma$ lower limits are still broadly consistent with the maximum model ($f_*=1$ for all galaxies) shown in the right-hand panels.

We also report, in Table \ref{table:galaxies_burst}, the stellar masses, formation times and redshifts for our 4 galaxies under the assumption of our instantaneous burst SFH model. This provides an approximate lower limit on the stellar masses and the ages of these galaxies. The masses are almost indistinguishable from the values reported in Table \ref{table:galaxies}, whereas the ages are slightly younger as expected.

\label{lastpage}
\end{document}